\documentclass[preprint,showpacs,epsfig,eqsecnum,aps]{revtex4}
\usepackage{graphicx}
\usepackage{epsfig}
\begin{document}
\title{New results for the two neutrino double beta decay in deformed nuclei
with angular momentum projected basis}

\author{A. A. Raduta$^{a),b)}$, A. Escuderos$^{c)}$, Amand Faessler$^{d)}$,
E. Moya de Guerra$^{c)}$ and P.Sarriguren$^{c)}$}

\address{$^{a)}$Institute of Physics and Nuclear Engineering, Bucharest, POBox MG6, Romania}
\address{
$^{b)}$Department of Theoretical Physics and Mathematics,Bucharest University, POBox MG11,
Romania}
\address{$^{c)}$Instituto de Estructura de la Materia, Consejo Superior de Investigationes Cientificas,
Serrano 123, E-28006 Madrid, Spain}
\address{$^{d)}$Institut fuer Theoretische Physik, Universitaet
Tuebingen, D-72076 Tuebingen, Germany}

\begin{abstract}
Four nuclei which are proved to be $2\nu\beta\beta$ emitters
( $^{76}$Ge, $^{82}$Se, $^{150}$Nd, $^{238}$U), and four suspected,
due to the corresponding Q-values,
to have this property ( $^{148}$Nd, $^{154}$Sm, $^{160}$Gd, $^{232}$Th),
were treated within a proton-neutron quasiparticle random phase approximation
(pnQRPA) with a projected spherical single particle basis. The advantage of
the present procedure over the ones
using a deformed Woods Saxon or Nilsson single particle basis is that the
actual pnQRPA states have a definite angular momentum while  all the others
provide
states having only K as a good quantum number.
The model Hamiltonian involves a mean field term yielding the projected single
particle states,
a pairing interaction for alike nucleons and a dipole-dipole proton-neutron
interaction in both the particle-hole (ph) and particle-particle (pp) channels.
The effect of nuclear deformation on the single beta strength distribution as well as on the
double beta Gamow-Teller transition amplitude (M$_{{\rm GT}}$) is analyzed.
The results are compared with the existent data and with the results from  a
different approach, in terms of the process half life T$_{1/2}$. The case of
different deformations for mother and daughter nuclei is also presented.
\end{abstract}

\maketitle

\section{Introduction}
\label{sec:level1}
One of the most exciting nuclear physics subject is that of double beta decay.
The interest is generated by the fact that in order to describe quantitatively
the decay rate one has to treat consistently the neutrino properties as well as the
nuclear structure features.
The process may take place in two distinct ways: a) by a $2\nu\beta\beta$
decay the initial nuclear system, the mother nucleus, is transformed in
  the final stable nuclear system, usually called
the daughter nucleus, two electrons and two anti-neutrinos
b) by the $0\nu\beta\beta$ process
the final state  does not involve any neutrino. The latter decay
mode is
especially interesting since one hopes that its discovery might provide a definite
answer to the question whether the neutrino is a Majorana or a Dirac particle.
The $0\nu\beta\beta$ decay is an extremely rare process and moreover it is hard
to distinguish the electrons emerging from the two processes. For some processes
there exists information about the low limits  of the process half-lives.
Combining this
information with the nuclear matrix  elements, some conclusions about the upper
limits of both neutrino effective mass and effective right-handedness of the
electroweak interaction was possible.
Unfortunately there are no reliable tests for the nuclear matrix elements involved
and therefore some indirect methods should be adopted.
It is worth mentioning that the  matrix elements which are responsible for
neutrinoless double
beta decay are similar to those needed for calculating the $2\nu\beta\beta$ decay rate,
for which there exists experimental data.
Due to this feature an indirect test for the matrix elements used for $0\nu\beta\beta$
is to use those m.e. which describe quantitatively the $2\nu\beta\beta$ decay.

For such reasons many theoreticians focused their efforts in describing
consistently the data for $2\nu\beta\beta$ decay. The contributions over
several decades have
been reviewed by many authors. Instead of enumerating the main steps achieved
toward improving the
theoretical description we advise the reader to consult few of the review works
\cite{Suh,Ver,PriRo,HaSt,Tomo,Fass,Kla1,Rad1}.

It is interesting to note that although none of the double beta emitters is a
spherical nucleus
most formalisms use a single particle spherical basis. More than 10 years ago,
two of us \cite{Rad2}
 proposed a
formalism to describe the process of two neutrinos double beta decay in a projected
spherical basis. It was for the first time that a pnQRPA approach for a
two body interaction in
the ph and pp channels with a deformed single particle basis was performed.
 Moreover, effects which are beyond the proton-neutron quasiparticle random
phase approximations (pnQRPA) have been accounted for by means of a boson expansion
procedure. A few years later the influence of nuclear deformation upon the contribution
of the spin-flip
configurations to the Gamow-Teller double beta transition amplitude, was studied
\cite{Rad3}.
In the meantime several papers have been  devoted to the extension of the
pnQRPA procedure to deformed nuclei, the applications being performed for
studying the single beta decay
properties as well as the double beta decay rates.
Thus,
pnQRPA approaches using as deformed single particle basis, Nilsson or deformed Woods Saxon
states have been formulated\cite{Hom,Sar,Eng}. Also a selfconsistent deformed
method was formulated where the single particle basis was obtained as eigenstates of a
deformed mean field obtained through a
Hartree-Fock treatment of a density dependent two body interaction
of Skyrme type \cite{Sar}.

The present investigation is, in fact, a continuation of the work from
 Ref.\cite{Rad2}. Therein the single particle energies were depending linearly on a
parameter which simulates the nuclear deformation.
By contrast, here the core  volume conservation constraint, ignored in the previous paper,
determines a nonlinear deformation dependence for single particle energies.
Of course, having  different single particle energies one expects that the pairing
properties and the double beta matrix elements are modified.
Another issue addressed in the present paper is whether considering different
deformations for the mother  and daughter nuclei, modifies significantly the
double beta transition amplitude (M$_{GT}$).
To be more specific, we recall that the standard pnQRPA approach including only
the
two body interaction in the particle-hole (ph) channel yields a M$_{GT}$ value much
larger than the experimental value extracted from the corresponding half-life.
Apparently, the desired $M_{GT}$ suppression might be obtained by a suitable choice
of the two body interaction in the particle-particle (pp) channel.
However, the fitted strength is close to the value where M$_{GT}$ cancels and moreover
close to the critical value where the pnQRPA breaks down.
It is obvious that increasing the deformation for the daughter nucleus the pnQRPA phonon state
is less correlated and therefore the pnQRPA breaking point is pushed toward
 larger
values. In this respect, one may say that the value of the pp interaction strength
which reproduces the experimental value for M$_{GT}$ becomes reliable, i.e. the
corresponding pnQRPA ground state of the daughter nucleus is stable
against adding anharmonic effects.

The formalism and results of the present paper will be presented according to the
following plan. In Section II a brief review of the projected spherical single
particle basis will be presented. Section III deals with the pnQRPA treatment of
a many body Hamiltonian which describes the nuclear states of the mother, daughter and intermediate
odd-odd nuclei, involved in the $2\nu\beta\beta$ process.
In Section IV, we discuss the results for eight double beta emitters:
$^{76}$Ge, $^{82}$Se, $^{148}$Nd $^{150}$Nd, $^{154}$Sm, $^{160}$Gd,
 $^{232}$Th, $^{238}$U for which the strength distribution for single $\beta^-$ and
 $\beta^+$ emission for mother and daughter nuclei,respectively,
 the M$_{GT}$ and half lives values for the double beta
 decay process are presented. A short summary and concluding remarks are given in Section V.

\section{Projected single particle basis}
\label{sec:level2}
In Ref. \cite{Rad4}, one of us, (A.A.R.), introduced  an angular momentum projected single particle basis which seems to be appropriate
for the description of the single particle motion in a 
deformed mean field generated by the particle-core interaction. This single particle basis
has been used to study the collective M1 states in deformed nuclei \cite{Rad44}
as well as
the rate of double beta process \cite{Rad2,Rad3}. Recently a new version
has been proposed where the deformation
dependence of single particle energies is nonlinear and therefore more realistic
\cite{Rad5,Rad6}.
In order to fix the necessary notations and to be self-contained, in the
present work we  describe briefly
the main ideas underlying the construction of the projected single
particle basis. Also some new properties for the projected basis are pointed out.

The single particle mean field is determined by a particle-core Hamiltonian: 
\begin{equation}
\tilde{H}=H_{sm}+H_{core}-
M\omega_0^2r^2\sum_{\lambda=0,2}\sum_{-\lambda\le\mu\le\lambda}
\alpha_{\lambda\mu}^*Y_{\lambda\mu}.
\label{hpc}
\end{equation}
where  $H_{sm}$ denotes the spherical shell model Hamiltonian while $H_{core}$ is
a harmonic quadrupole boson ($b^+_\mu$) Hamiltonian associated to a phenomenological core. 
The interaction of the two subsystems is accounted for by 
the third term of the above equation, written in terms of the shape coordinates $\alpha_{00}, \alpha_{2\mu}$.
The quadrupole shape coordinates are related to the quadrupole boson
 operators by the canonical transformation:
\begin{equation}
\alpha_{2\mu}=\frac{1}{k\sqrt{2}}(b^{\dagger}_{2\mu}+(-)^{\mu}b_{2,-\mu}),
\label{alpha2}
\end{equation}
where $k$ is an arbitrary C number. The monopole shape coordinate is to be 
determined from the volume conservation condition. In the quantized form, the result is:
\begin{equation}
\alpha_{00}=-\frac{1}{4k^2\sqrt{\pi}}\left[5+\sum_{\mu}(2b^{\dagger}_{\mu}b_\mu
+(b^{\dagger}_{\mu}b^{\dagger}_{-\mu}+b_{-\mu}b_{\mu})(-)^{\mu})\right].
\label{alpha0}
\end{equation}
Averaging $\tilde{H}$ on the eigenstates of $H_{sm}$, hereafter denoted by
$|nljm\rangle$, one obtains a deformed boson Hamiltonian whose ground state 
is, in the harmonic limit, described by a coherent state
\begin{equation}
{\Psi}_g=exp[d(b_{20}^+-b_{20})]|0\rangle_b,
\label{psig}
\end{equation}
with $|0\rangle_b$ standing for the vacuum state of the boson operators and $d$ a real parameter 
which simulates the nuclear deformation.
On the other hand, the average of $\tilde{H}$ on ${\Psi}_g$ is  similar to
the Nilsson Hamiltonian \cite{Nils}.
Due to these properties, it is expected that the best trial functions 
to generate, through projection, a spherical basis are:
\begin{equation}
{\Psi}^{pc}_{nlj}=|nljm\rangle{\Psi}_g.
\label{psipc}
\end{equation}
The upper index appearing in the l.h. side of the above equation suggests that the product function is associated 
to the particle-core system.
The projected states are obtained, in the usual manner, by acting
on these deformed states with the projection operator
\begin{equation}
P_{MK}^I=\frac{2I+1}{8\pi^2}\int{D_{MK}^I}^*(\Omega)\hat{R}(\Omega)
d\Omega .
\label{pjmk}
\end{equation}
We consider the subset of projected states :
\begin{equation}
\Phi_{nlj}^{IM}(d)={\cal N}_{nlj}^IP_{MI}^I[|nljI\rangle\Psi_g]\equiv
{\cal N}_{nlj}^I\Psi_{nlj}^{IM}(d) .
\label{phiim}
\end{equation}
which are orthonormalized.

The main properties of these projected spherical states are:
a) They are orthogonal with respect to I and M quantum numbers.
b) Although the projected states are associated to the particle-core system,
they can be used as a single particle basis. Indeed, when a matrix element of 
a particle like operator is calculated, 
the integration on the core collective coordinates is performed first,  which 
results in obtaining a final 
factorized expression: one factor carries the dependence on deformation and
one is a spherical shell model matrix element.
c) The connection between the nuclear deformation and the parameter $d$ 
entering the definition of the coherent state (2.4) is readily obtained by
 requiring that the strength of the particle-core 
quadrupole-quadrupole interaction 
be identical to the Nilsson deformed term of the mean field:
\begin{equation}
\frac{d}{k}=\sqrt{\frac{2\pi}{45}}(\Omega^2_{\perp}-\Omega^2_z).
\label{dpek}
\end{equation}
Here $\Omega_{\perp}$ and $\Omega_z$ denote the frequencies of Nilsson's mean 
field related to 
the deformation $\delta=\sqrt{45/16\pi}\beta$ by:
\begin{equation}
\Omega_{\perp}=(\frac{2+\delta}{2-\delta})^{1/3},~
\Omega_{z}=(\frac{2+\delta}{2-\delta})^{-2/3}.~
\label{omegaper}
\end{equation}
The constant $k$ was already defined by Eq.\ref{alpha2}. This is at our disposal
since the canonical property of the quoted transformation is satisfied for any value of
$k$.
The average of the particle-core Hamiltonian $H'=\tilde{H}-H_{core}$
on the projected spherical states defined by Eq.(2.7) has the expression
\begin{eqnarray}
\epsilon_{nlj}^I&=&\langle\Phi_{nlj}^{IM}(d)|H'|\Phi_{nlj}^{IM}(d)\rangle
=\epsilon_{nlj}-\hbar\omega_0(N+\frac{3}{2})C_{I0I}^{j2j}C_{1/201/2}^{j2j}
\frac{(\Omega^2_{\perp}-\Omega^2_z)}{3}\nonumber\\
&+&\hbar\omega_0(N+\frac{3}{2})\left[1+\frac{5}{2d^2}+\frac
{\sum_{J}(C^{jIJ}_{I-I0})^2I^{(1)}_J}
{\sum_{J}(C^{jIJ}_{I-I0})^2I^{(0)}_J}\right]
\frac{(\Omega^2_{\perp}-\Omega^2_z)^2}{90}.
\label{epsI}
\end{eqnarray}
Here we used the Condon-Shortley convention and notation for the
Clebsch Gordan coefficients
$C^{j_1 j_1 j}_{m_1 m_2 m}$.  $I_J^{(k)}$ stands for the following integral
\begin{equation}
I^{(i)}_J=\int^{1}_{0}P_J(x)[P_2(x)]^iexp[d^2P_2(x)]dx, ~i=0,1.
\label{IkJ}
\end{equation}
where $P_J(x)$ denotes the Legendre polynomial of rank J. 
It is worth mentioning that the norms for the core's projected states as well as the matrix elements of any boson operator 
on these projected states can be
fully determined once the overlap integrals defined in (\ref{IkJ}), are known
\cite{Rad6}.
Since the core contribution does not depend on the quantum numbers of the 
single particle energy level, it  produces a shift for all energies and therefore is omitted in 
Eq.(2.10). However, when the ground state energy variation
against deformation is studied, this term must be included.

The first term from (\ref{epsI}) is, of course, the single particle energy for
the spherical shell model state $|nljm\rangle$. The second term, linear in the
deformation parameter
{\it d}, is the only one considered in the previous works devoted to the double beta
decay of deformed nuclei within a projected spherical basis formalism. The third term from Eq.(\ref{epsI}) is determined
by the monopole-monopole particle-core coupling term after
implementing the volume conservation
condition. This term is the one responsible for the nonlinear deformation
dependence of $\epsilon^I_{nlj}$. The energies $\epsilon^I_{nlj}$  are represented
as function of the deformation parameter $d$, for the major shells with N equal to
3 and 4, in Fig. 1.
We remark that the energies shown in the above mentioned plot depend on
deformation in a different manner than those obtained in Ref. \cite{Rad4}.
Indeed, therein
they depend linearly on deformation, while here non-linear effects are
present. The difference between the two sets of energies is caused by the fact
 that here the volume conservation condition was used 
for the monopole shape coordinate, while in Ref.\cite{Rad4} this term is ignored.
The difference in the single particle energies is expected to cause significant effects on
the single and double beta transition probabilities. Actually, this is the main
motivation for the present investigation.

As shown in Fig. 1, the dependence of the new single particle energies on
deformation is similar to that shown by
the Nilsson model\cite{Nils}.

Although the energy levels are similar to  those of the Nilsson model, the 
quantum numbers in the two schemes 
are different. Indeed, here we generate from each j a multiplet of $(2j+1)$
states distinguished by the quantum number I, which plays the role of the Nilsson quantum number $\Omega$, runs from 1/2 to j and moreover the energies
corresponding to the quantum numbers K
and -K are equal to each other.
On the other hand, for a given I there are $2I+1$ degenerate sub-states while the Nilsson states are only double degenerate. 
As explained in Ref.\cite{Rad4}, the redundancy
problem can be solved by changing the normalization of the model functions:
\begin{equation}
\langle\Phi_{\alpha}^{I M}|\Phi_{\alpha}^{I M}\rangle=1 \Longrightarrow \sum_{M}\langle\Phi_{\alpha}^{IM}|\Phi_{\alpha}^{IM}\rangle=2.
\label{newnorm}
\end{equation}
Due to this weighting factor the particle density function is providing the
consistency result that the number of particles which can be
distributed on the (2I+1) sub-states is at most 2, which agrees with the
Nilsson model.
Here $\alpha$ stands for the set of shell model quantum numbers $nlj$.
Due to this normalization, the states $\Phi^{IM}_{\alpha}$ used to calculate the
matrix elements of a given operator should be multiplied with the weighting factor $\sqrt{2/(2I+1)}$.
The role of the core component is to induce a quadrupole deformation for the matrix elements of the operators acting on particle 
degrees of freedom. Indeed, for any such an operator the following factorization holds:
\begin{equation}
\langle \Phi^I_{nlj}||T_k||\Phi^{I^{\prime}}_{n^{\prime}l^{\prime}j^{\prime}}\rangle =
f^{n^{\prime}l^{\prime}j^{\prime}I^{\prime}}_{nljI}\langle nlj||T_k||
n^{\prime}l^{\prime}j^{\prime}\rangle.
\label{redTk}
\end{equation}
The factor $f$ carries the dependence on the deformation parameter $d$ while
the other factor is just the 
reduced matrix elements corresponding to the spherical shell model states. 
For details we advise the reader to consult Refs.\cite{Rad4,Rad6}.

Concluding, the projected single particle basis is defined by Eq. (\ref{phiim}).
Although these states are associated to a particle-core system, they can be used as a single particle basis due to the properties 
mentioned above.

Therefore,
the projected states might be thought of as eigenstates of an effective rotational invariant fermionic one-body 
Hamiltonian $H_{eff}$, with the corresponding  energies given by Eq.(\ref{epsI}).
\begin{equation}
H_{eff}\Phi^{IM}_{\alpha}=\epsilon^I_{\alpha}(d)\Phi^{IM}_{\alpha}.
\label{Heff}
\end{equation}
This definition should be supplemented by the request that the matrix elements of any operator between states $\Phi^{IM}_{\alpha}$ and 
$\Phi^{I^{\prime}M^{\prime}}_{\alpha^{\prime}}$, are given by Eq. (\ref{redTk}).
Due to these features, these states can be used as single particle basis to
treat many body Hamiltonians which involve one-body operators.
This is the case of Hamiltonians with two body separable forces. As a matter of fact,
such a type of Hamiltonian is used in the present paper.

According to our remark concerning the use of the projected spherical
states for describing the single particle motion, the average values
$\epsilon_{nlj}^I$ may be viewed as approximate expressions for the single
particle energies in deformed Nilsson orbits\cite{Nils}.
We may account for the deviations
from the exact eigenvalues by considering, 
later on, the exact matrix elements of the two body interaction when a specific
treatment of the many body system is applied.

Few words about the vibrational limit, $d\to 0$, for the projected basis are
necessary.
It can be proved that the following relations hold:
\begin{eqnarray}
\lim_{d\to 0}\Psi^{IM}_{nlj}&=&\delta_{I,j}|nljM\rangle|0\rangle_b, \nonumber\\
\lim_{d\to 0}\left({\cal N}^I_{nlj}\right)^{-1}&=&\delta_{I,j},\nonumber\\
\lim_{d\to 0}\langle \Phi^j_{nlj}||T_k||\Phi^{j^{\prime}}_{n^{\prime}l^{\prime}j^{\prime}}\rangle
&=&\langle nlj||T_k||n^{\prime}l^{\prime}j^{\prime}\rangle,
\nonumber\\
\lim_{d\to 0}\epsilon^j_{nlj}&=&\epsilon_{nlj}.
\label{limi}
\end{eqnarray}
Note that in the limit $d\to 0$, the norms of the states with $I\ne j$ are not
defined
while the limit of the I=j state, normalized to unity, is just the product
state $|nljM\rangle |0\rangle_b$. Indeed, the fourth equation (\ref{limi}) is
fulfilled
by neglecting a small quantity ($\frac{5}{8\pi k^2}$) caused by the zero point
motion term of the monopole-monopole particle-core interaction.
Although in the limit $d\to 0$ the norm of the states  $I\ne j$
is not defined, the limit of
$\Phi^{IM}_{nlj}$, with $I\ne j$, exists. However, the corresponding energies
are not identical to but very close to the
spherical shell model state energy.
\begin{equation}
\lim_{d\to 0}\epsilon^I_{nlj}=\epsilon_{nlj}+\hbar \omega_0(N+\frac{3}{2})
\left(\frac{5}{2}+\frac{1}{2}\left(j-I+\frac{1}{2}(1-(-)^{j-I})\right)\right)\frac{1}{4\pi k^2},\;j\ne I.
\end{equation}
  Indeed, this term should be compared with the spherical oscillator energy
 ($ (N+\frac{3}{2})\hbar \omega_0$) from $\epsilon_{nlj}$. Since the factor
 $\frac{1}{4\pi k^2}$ is very small (see Table 1) the correction of the
 shell model term is negligible.

Due to the properties mentioned above, we may state that in the
vibrational limit, $d\to 0$, the projected spherical basis goes to the spherical shell
model basis.

To complete our description of the projected single particle basis, we recall a
fundamental result obtained in Ref.\cite{Rad6}, concerning the product of two single
particle states which comprises a product of two core components.
Therein we have proved that the
matrix elements of a two body interaction corresponding to the present scheme are
very close to the matrix elements corresponding to spherical
states projected from  a  deformed product state with one factor as  a product of
two spherical single
  particle states, and a second factor consisting of a common collective
  core wave function. The small discrepancies of the two types of matrix elements
  could be washed out by using  slightly different strengths for the two body interaction in the two
  methods.

\section{The model Hamiltonian and its pnQRPA approach} 
\label{sec:level3} 

As we already stated, in the present work we are interested to describe
the Gamow-Teller two neutrino double beta decay of an even-even deformed nucleus.
In our treatment the Fermi transitions, contributing about 20\% and the
``forbidden''
transitions are ignored which is a reasonable approximation for
the two neutrino double beta decay in medium and heavy nuclei.
Customarily, the $2\nu\beta\beta$ process is conceived as two successive
single
$\beta^-$ transitions. The first transition connects the ground state of
the mother nucleus to a magnetic dipole state $1^+$ of the intermediate odd-odd nucleus which
subsequently decays to the ground state of the daughter nucleus.
Going beyond the pnQRPA procedure by means of the boson expansion procedure we
were able to consider the process leaving the final nucleus in an excited
collective state \cite{Rad7}. Such processes are not treated in the present paper.
The states, mentioned above, involved in the $2\nu\beta\beta$
 process are described by the following many body Hamiltonian:

\begin{eqnarray}
H=&&\sum\ \frac{2}{2I+1}(\epsilon_{\tau\alpha I}-
\lambda_{\tau\alpha})c^{\dagger}_{\tau\alpha IM}c_{\tau \alpha IM}-\sum\frac{G_{\tau}}{4}
P^{\dagger}_{\tau \alpha I}P_{\tau\alpha I'}\nonumber \\
 &+& 2\chi\sum\beta^-_{\mu}(pn)\beta^+_{-\mu}(p'n')(-)^{\mu}
 -2\chi_1\sum P^-_{1\mu}(pn)P^+_{-\mu}(p'n')(-)^{\mu}.
\label{Has}
\end{eqnarray}
The operator $c^{\dagger}_{\tau\alpha IM}(c_{\tau\alpha IM})$
creates (annihilates) a particle of type $\tau$ (=p,n)
in the state $\Phi^{IM}_{\alpha}$, when acting on the vacuum
state $|0\rangle$. In order to simplify the notations, hereafter the set of
quantum numbers $\alpha(=nlj)$ will be omitted. The two body interaction
consists of three terms, the pairing, the dipole-dipole particle hole (ph) and
the particle-particle (pp) interactions. The corresponding strengths are
denoted by $G_{\tau},\chi,\chi_1$, respectively. All of them are separable
interactions, with the factors defined by the following expressions:

\begin{eqnarray}
P^{\dagger}_{\tau I}&=&\sum_{M}\frac{2}{2I+1}c^{\dagger}_{\tau IM}
c^{\dagger}_{\widetilde{\tau IM}},\nonumber\\
\beta_{\mu}(pn)&=&\sum_{M,M'}\frac{\sqrt{2}}{{\hat I}}
\langle pIM|\sigma_{\mu}|n I'M'\rangle \frac{\sqrt{2}}{{\hat {I'}}}
c^{\dagger}_{pIM}c_{nI'M'},\nonumber\\
P^-_{1\mu}(pn)& = & \sum_{M,M'} \frac{\sqrt{2}}{{\hat I}}\langle pIM|\sigma_{\mu}|nI'M'\rangle \frac{\sqrt{2}}{{\hat {I'}}}
c^{\dagger}_{pIM}c^{\dagger}_{\widetilde{nI'M'}}.
\label{Psibeta}
\end{eqnarray}
The remaining operators from Eq.(\ref{Has}) can be obtained from the above defined operators
by hermitian conjugation.

The one body term and the pairing interaction terms are treated first through
the standard BCS formalism and consequently replaced by the quasiparticle
one body term $\sum_{\tau IM}E_{\tau}a^{\dagger}_{\tau IM}a_{\tau IM}$.
In terms of quasiparticle creation ($a^{\dagger}_{\tau IM}$) and annihilation
($a_{\tau IM}$) operators, related to the particle operators by means of the
Bogoliubov-Valatin transformation, the two body interaction terms, involved
in the model Hamiltonian, can be expressed just by replacing the
operators (3.2) by their quasiparticle images:

\begin{eqnarray}
\beta^-_{\mu}(k)&=&\sigma_kA^{\dagger}_{1\mu}(k)+\bar{\sigma}_kA_{1,-\mu}(k)(-)^{1-\mu}
+\eta_kB^{\dagger}_{1\mu}(k)-\bar{\sigma}_kB_{1,-\mu}(k)(-)^{1-\mu},\nonumber\\
\beta^+_{\mu}(k)&=&-\left[\bar{\sigma}_kA^{\dagger}_{1\mu}(k)+\sigma_kA_{1,-\mu}(k)(-)^{1-\mu}
-\bar{\eta}_kB^{\dagger}_{1\mu}(k)+\sigma_kB_{1,-\mu}(k)(-)^{1-\mu}\right],\nonumber\\
P^-_{1\mu}(k)&=&\eta_kA^{\dagger}_{1\mu}(k)-\bar{\eta}_kA_{1,-\mu}(k)(-)^{1-\mu}
-\sigma_kB^{\dagger}_{1\mu}(k)+\bar{\sigma}_kB_{1,-\mu}(k)(-)^{1-\mu}\nonumber,\\
P^+_{\mu}(k)&=&-\left[-\bar{\eta}_kA^{\dagger}_{1\mu}(k)+\eta_kA_{1,-\mu}(k)(-)^{1-\mu}
+\bar{\sigma}_kB^{\dagger}_{1\mu}(k)-\sigma_kB_{1,-\mu}(k)(-)^{1-\mu}\right].
\label{bbP1P}
\end{eqnarray}
In the above equations the argument ``k'' stands for the proton-neutron state (p,n).
Here, the usual notations for the dipole two quasiparticle and quasiparticle density operator have been used:

\begin{eqnarray}
A^{\dagger}_{1\mu}(pn)&=&\sum_{m_p,m_n}C^{I_p\; I_n \; 1}_{m_p \; m_n \; \mu}
a^{\dagger}_{pI_pm_p}a^{\dagger}_{n I_n m_n},\nonumber\\
B^{\dagger}_{1\mu}(pn)&=&\sum_{m_p,m_n}C^{I_p\; I_n \; 1}_{m_p \;-m_n \; \mu}
a^{\dagger}_{pI_pm_p}a_{n I_n m_n}(-)^{I_n-m_n}.
\label{A1B1}
\end{eqnarray}

The coefficients $\sigma$ and $\eta$ are simple expressions of the reduced matrix elements of the Pauli matrix $\sigma$ and U and V coefficients:
\begin{eqnarray}
\sigma_k=\frac{2}{\hat{1}\hat{I}_n}\langle I_p||\sigma||I_n\rangle U_{I_p}V_{I_n},\;\;
\bar{\sigma}_k=\frac{2}{\hat{1}\hat{I}_n}\langle I_p||\sigma||I_n\rangle V_{I_p}U_{I_n},\nonumber\\
\eta_k=\frac{2}{\hat{1}\hat{I}_n}\langle I_p||\sigma||I_n\rangle U_{I_p}U_{I_n},\;\;
\bar{\eta}_k=\frac{2}{\hat{1}\hat{I}_n}\langle I_p||\sigma||I_n\rangle V_{I_p}V_{I_n},
\label{sigeta}
\end{eqnarray}

The quasiparticle Hamiltonian is further treated within the proton-neutron random phase approximation (pnQRPA), i.e. one
determines the operator
\begin{equation}
\Gamma^{\dagger}_{1\mu}=\sum_{k}[X(k)A^{\dagger}_{1\mu}(k)-Y(k)A_{1,-\mu}(k)(-)^{1-\mu}],
\label{Gama}
\end{equation}

which satisfies the restrictions:

\begin{equation}
[\Gamma_{1\mu},\Gamma^{\dagger}_{1\mu'}]=\delta_{\mu,\mu'},\;\; [H_{qp},\Gamma^{\dagger}_{1\mu}]=\omega\Gamma^{\dagger}_{1\mu}.
\label{boscom}
\end{equation}

These operator equations yield a set of algebraic equations for the  X (usually called forward going) and Y (named back-going) amplitudes:

\begin{eqnarray}
\left(\matrix{\cal{A}&\cal{B}\cr -\cal{B}& -\cal{A}}\right)\left(\matrix{X \cr Y}\right)=\omega\left(\matrix{X \cr Y}\right),
\end{eqnarray}
\begin{equation}
\sum_{k}[|X(k)|^2-|Y(k)|^2]=1.
\label{rpaec}
\end{equation}

The pnQRPA matrices $\cal{ A}$ and $\cal{B}$ have analytical expressions:

\begin{eqnarray}
{\cal A}_{k,k'}&=&(E_p+E_n)\delta_{pp'}\delta_{nn'}
+2\chi(\sigma_k\sigma_{k'}+\bar{\sigma}_k\bar{\sigma}_{k'})
-2\chi_1(\eta_k\eta_{k'}+\bar{\eta}_k\bar{\eta}_{k'}),\nonumber\\
{\cal B}_{k,k'}&=&
2\chi(\bar{\sigma}_k\sigma_{k'}+\sigma_k\bar{\sigma}_{k'})
+2\chi_1(\bar{\eta}_k\eta_{k'}+\eta_k\bar{\eta}_{k'}).
\label{arobro}
\end{eqnarray}
All quantities involved in the pnQRPA matrices have been already defined.
Note that the proton and neutron quasiparticle energies are denoted in an
abbreviated manner by $E_p$ and $E_n$, respectively.

As can be seen from Eq. (\ref{Has}) the ph interaction is repulsive while the pp
interaction has an attractive character. Due to this feature, for a critical
value of $\chi_1$ the  lowest root of the pnQRPA equations may become imaginary.
Suppose that $\chi_1$ is smaller than its critical value and therefore all RPA solutions (i.e. $\omega$) are real numbers and ordered as:
\begin{equation}
\omega_1\le\omega_2\le...\le \omega_{N_s}.
\label{omordo}
\end{equation}
Here $N_s$ stands for the total number of the proton-neutron pair states whose
angular momenta can couple to $1^+$ and moreover their quantum numbers $n,l$ are the same.
Hereafter the phonon amplitudes $X$ and $Y$ will be accompanied by a lower
index ``$i$'' suggesting that they correspond to the energy $\omega_i$.

Since our single particle basis states depend on the deformation parameter $d$,
so do the pnQRPA energies and amplitudes.
The pnQRPA ground state (the vacuum state of the RPA phonon operator)
describes an even-even system which might be alternatively the mother or the daughter nucleus.
In the two cases the gauge and nuclear deformation properties are different
which results in determining distinct pnQRPA phonon operators acting on different vacua describing the mother and daughter ground states, respectively.
Therefore, one needs an additional index distinguishing the phonon operators of the mother and daughter nuclei.
The single  phonon states are defined by the equations:
\begin{equation}
|1_{j,k}\rangle=\Gamma^{\dagger}_{j,k}|0\rangle_j,\;j=i,f;\; k=1,2,...N_s.
\label{rpast}
\end{equation}
Here the indices $i$ and $f$ stand for initial (mother) and final (daughter) nuclei, respectively.
This equation defines two sets of non-orthogonal states describing the  
neighboring odd-odd nucleus. The states of the first set may be fed by a beta minus
decay of the ground state of the mother nucleus while the states of the second set are populated with a beta plus transition operator from the ground state of the daughter nucleus.

If the energy carried by leptons in the intermediate state is approximated by
the sum of the rest energy of the emitted electron and half the Q-value of
the double beta decay process
\begin{equation}
\Delta E=m_ec^2+\frac{1}{2}Q_{\beta\beta},
\label{DeltaE}
\end{equation} 
the reciprocal value of the $2\nu\beta\beta$ half life can be factorized as:
\begin{equation}
(T^{2\nu\beta\beta}_{1/2})^{-1}=F|M_{GT}(0^+_i\rightarrow 0^+_f)|^2,
\label{T1/2}
\end{equation}
where F is an integral on the phase space, independent of the nuclear structure,
 while M$_{GT}$ stands for the Gamow-Teller transition amplitude and has
 the expression \footnote{$^{*)}$Throughout this paper the Rose \cite{Rose} convention
 for the Wigner Eckart theorem is used.}:

\begin{equation}
M_{GT}=\sqrt{3}\sum_{kk'}\frac{_i\langle0||\beta^+_i||1_k\rangle_i 
\mbox{}_i\langle1_k|1_{k'}\rangle_f 
\mbox{}_f\langle1_{k'}||\beta^+_f||0\rangle_f}{E_k+\Delta E+E_{1^+}}.
\label{MGT}
\end{equation}
In the above equation, the denominator consists of three terms: a) $\Delta E$,
which was already defined, b) the average value of the k-th pnQRPA energy
normalized to the particular value corresponding to k=1, i.e.

\begin{equation}
E_k=\frac{1}{2}(\omega_{i,k}+\omega_{f,k})-\frac{1}{2}(\omega_{i,1}+\omega_{f,1}),
\label{Ek}
\end{equation}
and c) the experimental energy for the lowest $1^+$ state.
The indices carried by the transition operators indicate that they act in the
space spanned by the pnQRPA states associated to the initial (i) or final
(f) nucleus. Details about the overlap matrix of the single phonon states in the mother and daughter nuclei are given in Appendix A.

Before closing this section we would like to say a few words about what is
specific to our formalism.
As we mentioned before the pnQRPA matrices depend on the deformation
parameter and therefore the RPA energies and states depend on deformation.
Moreover, in the case that the mother and daughter nuclei are characterized by
different nuclear deformations the RPA output for the two nuclei are affected
differently by deformation. These features make the pnQRPA formalism build up
with a deformed single particle basis quite tedious.
Besides these difficulties, one should keep in mind the fact that the usual
approaches define the states from the intermediate odd-odd nucleus not as a
state of
angular momentum 1 but states of a definite K, i.e., $K=\pm1,0$.  Under such
circumstances from the pnQRPA states, the components of good angular momentum
are to be projected out. This operation is usually performed in an
approximative way,
(by transposing the result obtained in the intrinsic frame,  to the laboratory
frame of reference) which might be justified only in the strong coupling regime.
Unfortunately, the double beta emitters are only moderately deformed, which makes
the approximation validity, questionable.
Actually this is the reason why the answer to the
question of how much the results obtained with deformed single particle basis
differ from the ones obtained with projected many body RPA state, is not yet known.

By contrast, since our single particle states are projected spherical states,
the RPA formalism is fully identical to that one which is usually employed for spherical
nuclei. Since in the vibrational limit, ($d\to 0$), our basis goes to the spherical
shell model basis, one may say that the present formalism provides a unified description
of spherical and deformed nuclei.

\section{Numerical results} 
\label{sec:level4} 
The formalism described in the previous sections was applied to
eight  nuclei among which four are proved to be double beta emitters
($^{76}$Ge, $^{82}$Se, $^{150}$Nd, $^{238}$U) \cite{Bara} and four
suspected, due to the corresponding $Q_{\beta\beta}$ value, to have this
property.

The spherical shell model parameters are those given in Ref. \cite{Ring}, i.e.
\begin{equation}
\hbar\omega_0=41A^{-1/3},\;\; C=-2\hbar\omega_0\kappa,\;\; D=-\hbar\omega_0\mu.
\label{shellm}
\end{equation}
For the proton system, the pair of strength parameters ($\kappa,\mu$) takes the values
(0.08;0.) for $^{76}$Ge, $^{76}$Se, $^{82}$Se, $^{82}$Kr, (0.0637;0.6) for 
$^{148,150}$Nd, $^{148-154}$Sm, $^{154,160}$Gd,$^{160}$Dy, (0.0577;0.65)
for $^{232}$Th, $^{232,238}$U, $^{238}$Pu, while for the neutron systems of
the three groups of nuclei mentioned above, the values are
(0.08;0.), (0.0637,0.42),(0.0635;0.325), respectively.

The projected spherical single particle basis, used in our calculations,
depend on another two parameters, the deformation $\it {d}$ and  the
factor $\it {k}$ of the transformation (2.2) relating the boson operators with the
quadrupole collective coordinate, according to Eq. (\ref{alpha2}). These were fixed as follows. For the
lightest nuclei, Ge, Se, Kr, involved in the process of the double beta decay,
 the two parameters were taken so that the relative energies of the states
 $|1f \frac{7}{2}\frac{7}{2}\rangle$ and
$|1d\frac{5}{2}\frac{1}{2}\rangle$  as well as the lowest root of the pnQRPA
equations with a
QQ interaction included, reproduces the relative energy of $\Omega=\frac{7}{2}$ and
$\Omega=\frac{1}{2}$ Nilsson states, in the N=3 major shell, and the experimental value for the first collective $2^+$ state.
The $d$ and $k$ parameters for $^{154}$Sm and its double beta partner $^{154}$Gd
 were taken equal to those used in a previous publication \cite{Rad6}, to describe the M1 states of the mother nucleus.
As for the remaining nuclei considered in this paper, the corresponding $d$
and $k$ parameters
are the same as in Ref.\cite{Rad8,Rad9} where one of us (A. A. R. ) described phenomenologically the spectroscopic properties of the
major rotational bands.

The BCS calculations have been performed within a single particle space restricted so
that at least the states from the proton and neutron major open shells are included.
Although the single particle energies depend on deformations here we keep calling
major shell a set of states having the same quantum number $\it {n}$, according to
Eq.(2.7). This truncation criterion defines  an energy interval for single particle
states. Of course, due to the level crossing caused by deformation, also states from
the lower proton and upper neutron major shells, lying in the energy interval defined before,
are included in the single particle space. Since only the proton-neutron pair
of states
characterized by the same orbital angular momenta, participate in single beta
decay processes,
the single particle spaces for proton and neutron systems are to be  the same.
It is well understood that the corresponding energies for protons and neutrons are however
different from each others for heavy isotopes, due to the mass dependence of the single particle mean field
strength parameters (4.1).

Pairing strengths have been fixed so that the mass differences of the
neighboring even-even nuclei are reproduced. The results are listed in Table 1.
Their values may be interpolated by  a linear  function of 1/A both for the mother :
\begin{equation}
G_p=\frac{12.186}{A}+0.06931,\;\;G_n=\frac{8.2745}{A}+0.11266,
\label{mothG}
\end{equation}
and the daughter nuclei:
\begin{equation}
G_p=\frac{13.806}{A}+0.06765,\;\;G_n=\frac{8.1563}{A}+0.13455.
\label{daugG}
\end{equation}
Slight deviations from these rules are registered for $G_p$ of
$^{76}$Ge and $^{76}$Se and $G_n$ of  $^{76}$Ge and $^{238}$Pu.
It is interesting to note that although the single particle basis is different
from the ones currently used in the literature, the results for the interaction strength is
quite close to the standard ones. For example for $^{150}$Sm the above equations are equivalent
to $G_p=22.86/A,\;G_n=25.624/A$.

As for the proton-neutron two body interactions, their strengths were taken as
in Ref.\cite{Hom} although the
single particle basis used therein, is different from ours:
\begin{equation}
\chi=\frac{5.2}{A^{0.7}} \rm{MeV},\;\;\chi_1=\frac{0.58}{A^{0.7}}\rm{MeV}.
\label{chi}
\end{equation}
The A dependence for the ph interaction strength has been derived by fitting
the position of the GT resonance for $^{40}$Ca, $^{90}$Zr and $^{208}$Pb.
The strength $\chi_1$ has been fixed so that the beta decay half lives of the nuclei with
$Z\le 40$ are reproduced.
Certainly, the A dependence for $\chi$ and $\chi_1$
depends on the mass region to which the considered
nucleus belongs as well as on the single particle space. As a matter of fact
our results for $^{76}$Ge and $^{82}$Se show that larger values for the ph interaction
strength improve the agreement with experimental data. Moreover, our comparison
suggests that a  certain caution should be taken when the mass dependence given
by Eq. (\ref{chi})
is considered as in these  nucleibe $\chi$ and $\chi_1$ parameters cannot be fixed by
single beta decay half lives as it is usually done.
Once the parameters defining the model Hamiltonian are fixed, the pnQRPA
equations have been solved for the mother and daughter nuclei and the output
results have been used in connection with Eq.(\ref{MGT}) to calculate the GT amplitude of the
$2\nu\beta\beta$ process. In the next step, the equation (\ref{T1/2}) provides the
double beta half life. The phase space factor F does not depend on
the structure of the nuclear states and therefore we take it as given in
Ref.\cite{Suh,Klap3}. The values for F, used in the present paper, correspond to
$g_A=1.254$.
Results for M$_{GT}$ and half lives (T$_{1/2 }$) are given in Table 2. There, we
also
give the strength of the ph and pp interactions produced by Eq.(\ref{chi}).
For comparison we present also the available experimental data \cite{Bara,Kir,Man,Elli} as well as the
results of Ref. \cite{Klap4} for $T_{1/2}$.

\begin{table}[h!]
\begin{tabular}{|c||c|c|c|c|c|c|}
\hline 
Nucleus & d & k & G$_{\rm p}$ [MeV] & G$_{\rm n}$ [MeV] & $\chi$ [MeV] & g$_{\rm pp}$ \\
\hline \hline
$^{76}$Ge & 1.9 & 7.1 & 0.300 & 0.295 & 0.35 & 0.112 \\
$^{76}$Se & & & 0.295 & 0.285 & & \\ \hline
$^{82}$Se & 1.6 & 3.5 & 0.150 & 0.160 & 0.35 & 0.112 \\
$^{82}$Kr & & & 0.210 & 0.215 & & \\ \hline
$^{148}$Nd & 1.555 & 10.81 & 0.118 & 0.200 & 0.15733 & 0.11154 \\
$^{148}$Sm & & & 0.120 & 0.220 & & \\ \hline
$^{150}$Nd & 1.952 & 9.89 & 0.160 & 0.150 & 0.15586 & 0.11154 \\
$^{150}$Sm & & & 0.190 & 0.190 & & \\ \hline
$^{154}$Sm & 2.29 & 5.58 & 0.190 & 0.134 & 0.15302 & 0.11154 \\
$^{154}$Gd & & & 0.145 & 0.138 & & \\ \hline
$^{160}$Gd & 2.714 & 4.384 & 0.160 & 0.155 & 0.14898 & 0.11154 \\
$^{160}$Dy & & & 0.155 & 0.160 & & \\ \hline
$^{232}$Th & 2.51 & 4.427 & 0.120 & 0.183 & 0.11486 & 0.11154 \\
$^{232}$U & & & 0.090 & 0.225 & & \\ \hline
$^{238}$U & 2.62 & 4.224 & 0.130 & 0.165 & 0.080 & 0.11154 \\
$^{238}$Pu & & & 0.165 & 0.235 & & \\
\hline
\end{tabular}
\caption{ The pairing and Gamow Teller interactions strength are given
in units of MeV. The ratio of the two dipole interaction
( particle-hole and particle-particle) strengths, denoted by $g_{pp}$ is 
given.
The list of the deformation parameter $d$ and the factor $k$ of the transformation
(2.2) are  also presented.
The manner in which these parameters were fixed is explained in the text.}
\end{table}

\begin{table}[h!]
\begin{tabular}{|c||c|c|c|c|c|c|}
\hline 
Nucleus & $\chi$ & g$_{\rm pp}$ & M$_{\rm GT}$ & \multicolumn{3}{c|}{T
$_{1/2}$ [yr]} \\ \cline{5-7}
        &        &              &              & present &exp.& Ref.\cite{Klap4}\\
\hline \hline
$^{76}$Ge $\to ^{76}$Se & 0.35 & 0.112 & 0.222 & $5.9 \cdot 10^{20}$ &
 9.2$^{+0.7}_{-0.4}\cdot 10^{20}$ $^{\rm a)}$&$2.61 \cdot 10^{20}$ \\
 & 0.35 & 0.112 & 0.149 $^{\rm c)}$ & $1.32 \cdot 10^{21}$ &
 1.1$^{+0.6}_{-0.3}\cdot 10^{21}$ $^{\rm b)}$&\\
                        &0.25&0.11154&0.270&$4.05 \cdot 10^{20}$&  & \\
 \hline
$^{82}$Se $\to ^{82}$Kr & 0.35 & 0.112 & 0.096 & $0.963 \cdot 10^{20}$ &$1.1^{+0.8}_{-0.3}\cdot 10^{20}$ $ ^{\rm d)}$ &$0.848 \cdot 10^{20}$
\\
                        &  0.16& 0.108  & 0.135 & $0.49 \cdot 10^{20}$&$1.0\pm0.4 \cdot 10^{20}$ $ ^{\rm e)}$  &  \\
                        &       &       &       &                     &$1.3\pm0.05 \cdot 10^{20}$ $ ^{\rm f)}$ &\\
\hline
$^{148}$Nd $\to ^{148}$Sm & 0.157 & 0.112 & 0.392 & $2.327 \cdot 10^{19}$ & & $1.19\cdot 10^{21}$  \\
\hline
$^{150}$Nd $\to ^{150}$Sm & 0.156 & 0.11154 & 0.350 & $2.630 \cdot 10^{17}$ &
 $\ge 1.8 \cdot 10^{19}$ $^{\rm g)}$&$1.66 \cdot 10^{19}$ \\
 & 0.156 & 1.50 & 0.040 & $1.98 \cdot 10^{19}$ & & \\
 \hline
$^{154}$Sm $\to ^{154}$Gd & 0.153 & 0.11154 & 0.327 & $8.760 \cdot 10^{20}$ & &
 $1.49 \cdot 10^{22}$  \\
 \hline
$^{160}$Gd $\to ^{160}$Dy & 0.149 & 0.11154 & 0.170 & $2.013 \cdot 10^{20}$ &
& $2.81 \cdot 10^{21}$  \\ \hline
$^{232}$Th $\to ^{232}$U & 0.11486 & 0.11154 & 0.123 & $4.240 \cdot 10^{21}$ &
& $4.03 \cdot 10^{21}$  \\ \hline
$^{238}$U $\to ^{238}$Pu & 0.080 & 0.112 & 0.166 & $2.375 \cdot 10^{21}$ &
 $(2.0 \pm 0.6)\cdot 10^{21}$ $^{\rm h)}$& $0.914 \cdot 10^{21}$ \\
                         &0.11282&0.11154&0.139&$3.340 \cdot 10^{21}$&  &\\
                         &0.08   &0.0    & 0.171&$2.249 \cdot 10^{21}$& &\\
\hline
\end{tabular}
\newline \newline
\caption{The Gamow-Teller amplitude for $2\nu\beta\beta$ decay, in units of
MeV$^{-1}$
and the corresponding  half life ($T_{1/2}$) are listed for several
ground to ground
transitions. The experimental half lives for the transitions
$^{76}$Ge$\to ^{76}$Se ($^{\rm a)}$Ref.\cite{Avi},$^{\rm b)}$Ref. \cite{Mil} ),
$^{82}$Se$\to ^{82}$Kr
($^{\rm d)}$ Ref. \cite{Elli}, $^{\rm e)}$ Ref.\cite{Man},
$^{\rm f)}$ Ref.\cite{Kir}),
 $^{150}$Nd $\to ^{150}$Sm ($^{\rm g)}$Ref.\cite{Kli})
 and
$^{238}$U $\to ^{238}$Pu  ($^{\rm h)}$ Refs. \cite{Tret,Bara,Vog,Turk}) are also given.
In the last column the results from
Ref.\cite{Klap4}, are given. The parameters $\chi$ and $g_{pp}$ are also given.
$^{c)}$For these two cases the mother and daughter nuclei have different
deformations, namely $d_i=1.6$ and $d_f=1.9.$ The parameters $\chi$ and
$g_{pp}$ are listed in the second and third columns.}
\end{table}

Before discussing in extenso the results from Table 2, it is instructive to
show the results concerning the single beta decay properties of the
mother and daughter nuclei. Thus, in Figs. 2-9 the
beta minus strength of the mother and the beta plus streghts of the daughter
nuclei, folded with a gaussian having the width equal to 1 MeV, are plotted as
function of energy.
For the lightest two nuclei, the experimental data are also presented \cite{Mad,Hel}.

For pedagogical reasons, for these two nuclei
two different ${\it ph}$ interaction strengths are alternatively used. In this way
one clearly sees that increasing the strength $\chi$, the transition strength
is moved to the higher state. Thus, although the peak positions remain the
same, since the BCS data are not changed, the first peak looses height
while the second one is augmented. The agreement with experimental data is
reasonably good. In both mother nuclei the center of the GT resonance is a little
bit shifted, backward for $^{82}$Se and forward for $^{76}$Ge. Due to the
fragmentation effect caused by the nuclear deformation, the theoretical result
for the GT resonance has a shorter, otherwise broader peak than the
experimental one. In order to see that a broad peak in the folded beta
strength plot means, indeed, a fragmentation of the strength
distributed among several pnQRPA states, we show the unfolded strength for
$^{82}$Se and $^{232}$Th isotopes in Figs 10 and 11, respectively. For example, in the case of
$^{82}$Se, the folded strength
exhibits a first fat peak which has  a very short maximum before and a
``shoulder'' on the descending part (see the middle panels of Figs.3).
From Fig.10  one sees that to these
details correspond pnQRPA states which carry a strength represented by sticks
which dominate the grass spread around.

The total strength of the GT resonance is about the same as the corresponding experimental data.
However the fragmentation is causing a broad resonance which results in having
a shorter peak.
This fact might raise the question whether the deformation considered in the present paper is too large.
Indeed, the neutron system of $^{82}$Se is almost spherical since N (the neutron number) is close to a magic value.
Due to this feature we repeated the calculations with a very small ${\it d}$(=0.2) which
is close to the spherical limit.
As seen from the right panel the height of GT resonance corresponding to the
new deformation
is close to the experimental result. Since we kept the same parameter for the single particle mean field parameters,
e.g for the parameter ${\it k}$ defined in Eq. (2.2), the theoretical curve is shifted by about 1 MeV with respect to the experimental data.
Of course, the position of the GT  resonance depends also on the pairing strengths.
The parameters mentioned above where kept the same as for the initial deformation case (d=1.9), in order to
judge, by comparison, the effect coming from the nuclear deformation.
In conclusion, going from deformed to spherical single particle basis the GT resonance peak
    is getting higher and the width narrower.
    Actually when the calculated strength distribution is compared with the
    experimental data
    one has to restrict the discussion only to the position of the GT centroid
    and the total
    strength, since there is no experimental information about the resonance width.
    The narrow whith  seen, however, in Figs. 2 and 3 is caused by folding
    a single number indicating the total $\beta^-$ strength for the GT resonance
    which has the centroid at a given energy, with a Gaussian having the
    width equal to 1 MeV.

From the folded beta minus strengths graphs, we see that for heavier double beta
emitters there exists a small peak lying above the GT resonance.
\begin{table}[h!]
\begin{tabular}{|c||c|c|c|c|c|c|}
\hline 
Nucleus & \multicolumn{2}{c|}{1st peak} & \multicolumn{2}{c|}{2nd peak}
& \multicolumn{2}{c|}{3rd peak}\\ \cline{2-7}
 & Transition & Strength & Transition & Strength & Transition & Strength \\ \hline \hline
$^{76}$Ge & $\nu (4d \frac{5}{2} \frac{3}{2}) \to \pi (4d \frac{5}{2}\frac{1}{2})$ &
 1.084&
 $\nu(4d \frac{3}{2}\frac{1}{2}) \to \pi(4d \frac{5}{2}\frac{3}{2})$ & 1.740
 &$\nu(3f \frac{7}{2}\frac{5}{2}) \to \pi(3f \frac{5}{2}\frac{5}{2})$&1.718 \\
 & & & $\nu(3f \frac{7}{2}\frac{1}{2}) \to \pi(3f \frac{5}{2}\frac{3}{2})$ & 3.371& & \\
\hline
$^{82}$Se & $\nu(3f \frac{5}{2}\frac{5}{2}) \to \pi(3f \frac{7}{2}\frac{3}{2})$ & 1.554 &
 $\nu (3p \frac{3}{2}\frac{3}{2}) \to \pi(3p \frac{3}{2}\frac{1}{2})$ & 4.501 &
 $\nu(3p\frac{3}{2}\frac{1}{2}\to \pi(3p\frac{3}{2}\frac{3}{2})$ &1.686 \\
&&&&&$\nu(3f \frac{7}{2}\frac{3}{2}) \to \pi(3f \frac{5}{2}\frac{3}{2})$ & 3.310 \\
\hline
$^{148}$Nd & $\nu(4g \frac{7}{2}\frac{7}{2}) \to \pi(4g \frac{9}{2}\frac{5}{2})$ & 1.077
 &$\nu(4g \frac{7}{2}\frac{3}{2}) \to \pi(4g \frac{9}{2}\frac{5}{2})$&  1.453& & \\
 & $\nu(4g \frac{7}{2}\frac{5}{2}) \to \pi(4g \frac{9}{2}\frac{7}{2})$ & 1.141 &
$\nu(4g \frac{9}{2}\frac{5}{2}) \to \pi(4g \frac{7}{2}\frac{3}{2})$&12.437& & \\
 & & &$\nu(4g \frac{9}{2}\frac{5}{2}) \to \pi(4g \frac{7}{2}\frac{7}{2})$ & 1.015 & & \\
 \hline
$^{150}$Nd & $\nu(4g \frac{7}{2}\frac{7}{2}) \to \pi(4g \frac{9}{2}\frac{5}{2})$ & 1.901 &
 $\nu(5h \frac{11}{2}\frac{3}{2}) \to \pi(5h \frac{9}{2}\frac{3}{2})$ & 1.246&
 $\nu(4d \frac{5}{2}\frac{1}{2}) \to \pi(4d \frac{3}{2}\frac{3}{2})$ &5.386 \\
& & &$\nu(4g \frac{7}{2}\frac{5}{2}) \to \pi(4g\frac{9}{2}\frac{3}{2})$ &1.178&
 $\nu(4g \frac{9}{2}\frac{3}{2}) \to \pi(4g \frac{7}{2}\frac{3}{2})$ & 2.370\\
 & & & $\nu(4g \frac{9}{2}\frac{5}{2}) \to \pi(4g \frac{7}{2}\frac{3}{2})$ &
 1.087&& \\
 & & & $\nu(4g \frac{9}{2}\frac{3}{2}) \to \pi(4g \frac{9}{2}\frac{5}{2})$ & 7.647&&\\
\hline
$^{154}$Sm & $\nu(5f \frac{7}{2}\frac{3}{2}) \to \pi(5f \frac{7}{2}\frac{5}{2})$ & 1.220 &
 $\nu(4g \frac{7}{2}\frac{1}{2}) \to \pi(4g \frac{7}{2}\frac{3}{2})$ & 4.214&
$ \nu(4g \frac{9}{2}\frac{5}{2}) \to \pi(4g \frac{7}{2}\frac{5}{2})$ & 1.537\\
 & $\nu(6i \frac{13}{2}\frac{7}{2}) \to \pi(6i \frac{13}{2}\frac{9}{2})$ & 1.031 &
 $\nu(4d \frac{3}{2}\frac{1}{2}) \to \pi(4d \frac{3}{2}\frac{3}{2})$ & 2.088&
$ \nu(4d \frac{5}{2}\frac{3}{2}) \to \pi(4d \frac{3}{2}\frac{3}{2})$ &4.380 \\
 & $\nu(4g \frac{7}{2}\frac{3}{2}) \to \pi(4g \frac{7}{2}\frac{1}{2})$ & 3.362 &
 & && \\
\hline
$^{160}$Gd 
 & $\nu (6i \frac{13}{2}\frac{5}{2}) \to \pi (6i \frac{11}{2}\frac{3}{2})$ & 2.222 &
 $\nu (4d \frac{5}{2}\frac{3}{2}) \to \pi (4d \frac{5}{2}\frac{1}{2})$ & 2.551&
$\nu (4d \frac{5}{2}\frac{3}{2}) \to \pi (4d \frac{3}{2}\frac{3}{2})$ &1.028  \\
&$\nu (4g \frac{9}{2}\frac{7}{2}) \to \pi (4g \frac{9}{2}\frac{9}{2})$ & 1.622 &
$\nu (5h \frac{11}{2}\frac{1}{2}) \to \pi (5h \frac{9}{2}\frac{3}{2})$ & 1.362&
$\nu (4d \frac{3}{2}\frac{1}{2}) \to \pi(4d \frac{3}{2}\frac{3}{2})$ &1.204  \\
& & & $\nu (6i\frac{13}{2}\frac{1}{2}) \to \pi(6i\frac{11}{2}\frac{3}{2})$ &9.922 &&\\
\hline
\end{tabular}
\caption{ Here are listed the strengths carried by the pnQRPA states
contributing to
the
first, second and third (if any) peaks from the upper panels of Figs.2-9.
On the left side of these numbers the 2qp configurations closest
in energy to the corresponding pnQRPA states, are given. Actually this is the dominant
configuration of the chosen pn phonon state.}
\end{table}
\clearpage
\begin{table}[!]
\begin{tabular}{|c||c|c|c|c|c|c|}
\hline 
Nucleus & \multicolumn{2}{c|}{1st peak} & \multicolumn{2}{c|}{2nd peak} & 
\multicolumn{2}{c|}{3rd peak} \\ \cline{2-7}
 & Transition & Strength & Transition & Strength & Transition & Strength \\ \hline \hline
$^{232}$Th 
 & $\nu(5h \frac{9}{2}\frac{7}{2}) \to \pi(5h \frac{9}{2}\frac{9}{2})$ & 1.810 &
 $\nu(4d \frac{3}{2}\frac{3}{2}) \to \pi(4d \frac{3}{2}\frac{1}{2})$ & 1.900 &
 $\nu(5p \frac{3}{2}\frac{1}{2}) \to \pi(5p \frac{3}{2}\frac{3}{2})$ & 2.548 \\
 &$\nu(5f \frac{5}{2}\frac{1}{2}) \to \pi(5f \frac{5}{2}\frac{3}{2})$ &1.569&
 $\nu(5h \frac{9}{2}\frac{3}{2}) \to \pi(5h \frac{11}{2}\frac{5}{2})$ & 9.913 & & \\
 & & & $\nu(5h \frac{11}{2}\frac{3}{2}) \to \pi(5h \frac{9}{2}\frac{5}{2})$ &
 7.157 & & \\
 & & & $\nu(5h \frac{11}{2}\frac{5}{2}) \to \pi(5h \frac{9}{2}\frac{3}{2})$ &
 2.190 & & \\
 & & & $\nu(6i \frac{13}{2}\frac{5}{2}) \to \pi(6i \frac{11}{2}\frac{7}{2})$ &
  2.392 & & \\
 & & & $\nu(5h \frac{11}{2}\frac{7}{2}) \to \pi(5f \frac{11}{2}\frac{5}{2})$ &
 8.709 & & \\
 \hline
$^{238}$U & $\nu(6g \frac{9}{2}\frac{3}{2}) \to \pi(6g \frac{7}{2}\frac{1}{2})$ & 2.253 &
 $\nu(6i \frac{13}{2}\frac{3}{2}) \to \pi(6i \frac{13}{2}\frac{1}{2})$ & 4.259 &
 $\nu(5p \frac{3}{2}\frac{1}{2}) \to \pi(5p \frac{3}{2}\frac{1}{2})$ & 1.397 \\
&$\nu(6i \frac{13}{2}\frac{1}{2}) \to \pi(6i \frac{13}{2}\frac{1}{2})$ &1.079&
 $\nu(6i \frac{13}{2}\frac{3}{2}) \to \pi(6i \frac{11}{2}\frac{5}{2})$ &2.868&&\\
& $\nu(5f \frac{5}{2}\frac{1}{2}) \to \pi(5f \frac{5}{2}\frac{3}{2})$ &1.888 &
$\nu(5h \frac{9}{2}\frac{3}{2}) \to \pi(5h \frac{9}{2}\frac{1}{2})$ & 1.112 & & \\
& & & $\nu(5h \frac{9}{2}\frac{1}{2}) \to \pi(5h \frac{9}{2}\frac{3}{2})$ & 6.201 & &\\
& & & $\nu(5h \frac{9}{2}\frac{1}{2}) \to \pi(5h \frac{9}{2}\frac{1}{2})$ & 2.912 & & \\
 & & & $\nu(4d \frac{5}{2}\frac{5}{2}) \to \pi(4d \frac{5}{2}\frac{3}{2})$ & 2.019 & & \\
 & & & $\nu(6g \frac{9}{2}\frac{1}{2}) \to \pi(6g \frac{7}{2}\frac{3}{2})$ & 3.662 & & \\
 & & & $\nu(5f \frac{7}{2}\frac{1}{2}) \to \pi(5f \frac{5}{2}\frac{3}{2})$ & 1.222 & & \\
\hline
\end{tabular}
\caption{ Continuation of Table III.}
\end{table}
\clearpage
\begin{table}[!]
\begin{tabular}{|c||c|c|c|c|c|c|}
\hline 
Nucleus & \multicolumn{2}{c|}{1st peak} & \multicolumn{2}{c|}{2nd peak}
& \multicolumn{2}{c|}{3rd peak} \\ \cline{2-7}
 & pnQRPA energy & Strength & pnQRPA energy & Strength& pnQRPA energy& Strength \\ \hline \hline
$^{76}$Ge & 7.033 & 1.084 & 10.850 & 1.740&12.602&1.718 \\
 & & & 11.605 & 3.371&& \\
\hline
$^{82}$Se & 6.939 & 1.554 & 10.920 & 4.501&11.701&1.686 \\
 & & & & &12.291&3.310 \\
\hline
$^{148}$Nd & 9.397 & 1.077 &12.028&1.453 && \\
 & 10.047 & 1.141 &12.269&12.437& & \\
 &  & &12.429&1.015& & \\
 \hline
$^{150}$Nd & 8.600 & 1.901 & 11.263 & 1.246& 12.939 & 5.386 \\
 & & & 11.531 & 1.178&13.217 &2.370 \\
 & & & 12.281 & 1.087&& \\
 & & & 12.597 & 7.647&& \\
\hline
$^{154}$Sm & 10.475 & 1.220 & 11.986 & 4.214&13.189&1.537 \\
 & 11.047 & 1.031 & 12.696 & 2.088 & 13.303 & 4.380 \\
 & 11.434 & 3.362 & & & & \\
\hline
$^{160}$Gd & 10.748 & 2.222 & 12.457 & 2.551&15.334&1.028 \\
 & 11.163 & 1.622 & 12.857 & 1.362&15.850&1.204 \\
 &  &  & 13.369 & 9.927& & \\
\hline
\end{tabular}
\caption{ The energies of the pnQRPA states which give the largest strength
contributions to  the peaks in Figs. 2-9, upper panels.
The carried strengths are also given.}
\end{table}

\clearpage
\begin{table}[!]
\begin{tabular}{|c||c|c|c|c|c|c|}
\hline 
Nucleus & \multicolumn{2}{c|}{1st peak} & \multicolumn{2}{c|}{2nd peak} & 
\multicolumn{2}{c|}{3rd peak} \\ \cline{2-7}
 & pnQRPA energy & Strength & pnQRPA energy & Strength & pnQRPA energy & Strength \\ \hline \hline
$^{232}$Th & 12.895 & 1.810 & 15.622 & 1.900 &
 19.559 & 2.548 \\
 &13.578 & 1.569& 15.952 & 9.913 & & \\
 & & & 16.367 &7.157 & & \\
 & & & 16.593 & 2.190 & & \\
 & & & 16.731 & 2.392 & & \\
 & & & 16.942 & 8.709 & & \\
\hline
$^{238}$U & 13.641 & 2.253 & 16.079 & 4.259 & 21.137 & 1.397 \\
 & 14.792 & 1.079 & 16.306 & 2.868 & & \\
 & 15.219 & 1.888 & 16.452 & 1.112 & & \\
 & & & 16.559 & 6.201 & & \\
 & & & 16.613 & 2.912 & & \\
 & & & 16.831 & 2.019 & & \\
 & & & 17.391 & 3.662 & & \\
 & & & 18.232 & 1.222 & & \\
\hline
\end{tabular}
\caption{ Continuation of Table V.}
\end{table}
\clearpage
\begin{table}[!]
\begin{tabular}{|c||c|c|c|c|c|c|}
\hline 
Nucleus & \multicolumn{2}{c|}{1st peak} & \multicolumn{2}{c|}{2nd peak} & 
\multicolumn{2}{c|}{3rd peak} \\ \cline{2-7}
 & Transition & Strength & Transition & Strength & Transition & Strength \\ \hline \hline
$^{76}$Se
 & $\pi(3f \frac{7}{2}\frac{7}{2}) \to \nu(3f \frac{5}{2}\frac{5}{2})$ & 0.313 &
 $\pi(3f \frac{7}{2}\frac{5}{2}) \to \nu(3f \frac{5}{2}\frac{5}{2})$ & 0.116 &
$\pi(3f\frac{5}{2}\frac{5}{2})\to \nu (3f \frac{7}{2}\frac{3}{2})$&0.013 \\
 & & & $\pi(3f \frac{7}{2}\frac{5}{2}) \to \nu(3f \frac{5}{2}\frac{3}{2})$ & 0.027  & &
 \\
 &&& $\pi(3f \frac{7}{2}\frac{3}{2}) \to \nu(3f \frac{5}{2}\frac{5}{2})$ & 0.021 & &\\
\hline
$^{82}$Kr
 & $\pi(3f \frac{7}{2}\frac{7}{2}) \to \nu(3f \frac{5}{2}\frac{5}{2})$ & 0.135 &
 $\pi(3p \frac{3}{2}\frac{3}{2}) \to \nu(3p \frac{1}{2}\frac{1}{2})$ & 0.118 &
 $\pi(3f \frac{7}{2}\frac{3}{2}) \to \nu(3f \frac{5}{2}\frac{5}{2})$ & 0.019\\
\hline
$^{148}$Sm
 & $\pi(5h \frac{11}{2}\frac{3}{2}) \to \nu(5h \frac{9}{2}\frac{1}{2})$ & 0.139 &
 $\pi(4d \frac{5}{2}\frac{1}{2}) \to \nu(4d \frac{3}{2}\frac{1}{2})$ & 0.037  &
 $\pi(4g\frac{9}{2}\frac{7}{2})\to \nu(4g \frac{7}{2}\frac{5}{2})$&0.011
\\
 & $\pi(5h \frac{11}{2}\frac{3}{2}) \to \nu(5h \frac{9}{2}\frac{3}{2})$ & 0.151 &
 $\pi(4g\frac{9}{2}\frac{9}{2})\to \nu (4g \frac{7}{2}\frac{7}{2})$ &0.017 &
 $\pi(4g \frac{7}{2}\frac{3}{2}) \to \nu(4g \frac{9}{2}\frac{5}{2})$ & 0.012\\
 &$\pi(5h \frac{11}{2}\frac{3}{2})\to \nu (5h \frac{9}{2}\frac{3}{2})$&0.165 & & & \\
\hline
$^{150}$Sm
 & $\pi(5h \frac{11}{2}\frac{3}{2}) \to \nu(5h \frac{9}{2}\frac{1}{2})$ & 0.227 &
$\pi(5h \frac{9}{2}\frac{1}{2})\to \nu (5h\frac{9}{2}\frac{3}{2})$ &0.012 &
&  \\
& $\pi(5h \frac{11}{2}\frac{1}{2}) \to \nu(5h \frac{9}{2}\frac{3}{2})$ & 0.134 &
$\pi (4d\frac{5}{2}\frac{1}{2})\to \nu (4d \frac{3}{2}\frac{1}{2})$&0.018 & &  \\
&$\pi (5h\frac{11}{2}\frac{3}{2})\to \nu(5h\frac{9}{2}\frac{3}{2})$&0.103 &&& &\\
 \hline
$^{154}$Gd
 & $\pi(4d \frac{5}{2}\frac{3}{2}) \to \nu(4d \frac{3}{2}\frac{3}{2})$ & 0.081 &
 $\pi(5h \frac{11}{2}\frac{3}{2}) \to \nu(5h \frac{9}{2}\frac{3}{2})$ & 0.061 &
 $\pi(5h \frac{11}{2}\frac{3}{2}) \to \nu(5h \frac{9}{2}\frac{1}{2})$  &0.039  \\
 & $\pi(5h \frac{11}{2}\frac{5}{2}) \to \nu(5h \frac{9}{2}\frac{5}{2})$ & 0.140 &
$\pi(5h \frac{11}{2}\frac{1}{2}) \to \nu(5h \frac{9}{2}\frac{3}{2})$  &0.046 &
$\pi(4d \frac{5}{2}\frac{1}{2}) \to \nu(4d \frac{3}{2}\frac{3}{2})$  &0.019\\
&     &  &  &  &
$\pi(4g \frac{9}{2}\frac{7}{2}) \to \nu(4g \frac{7}{2}\frac{7}{2})$&0.016\\
\hline
$^{160}$Dy
 & $\pi(4d \frac{5}{2}\frac{3}{2}) \to \nu(4d \frac{3}{2}\frac{3}{2})$ & 0.050 &
 $\pi(5h\frac{9}{2}\frac{1}{2}) \to \nu (5h \frac{9}{2}\frac{3}{2})$ &0.067&
 $\pi (6i \frac{13}{2}\frac{1}{2})\to \nu (6i \frac{11}{2}\frac{3}{2})$&0.063  \\
 & $\pi(5h \frac{11}{2}\frac{5}{2}) \to \nu(5h \frac{9}{2}\frac{5}{2})$ & 0.076
 &$\pi(4g\frac{9}{2}\frac{9}{2}) \to \nu(5h \frac{7}{2}\frac{7}{2})$ &0.113&
$\pi(4d\frac{5}{2}\frac{1}{2}) \to \nu(4d \frac{3}{2}\frac{3}{2})$ &0.051 \\
&$\pi (5h \frac{9}{2}\frac{3}{2})\to \nu (6i \frac{9}{2}\frac{5}{2})$&0.365&
 $\pi(6i \frac{13}{2}\frac{3}{2}) \to \nu(6i \frac{11}{2}\frac{1}{2})$ & 0.059 &
$\pi(5h \frac{11}{2}\frac{3}{2})\to \nu (5h \frac{9}{2}\frac{3}{2})$&0.054\\
&&&
$ \pi (5h \frac{11}{2}\frac{5}{2})\to \nu (5h \frac{9}{2}\frac{3}{2})$&0.073& &\\
\hline
$^{232}$U
 & $\pi(6i \frac{11}{2}\frac{7}{2}) \to \nu(6i \frac{13}{2}\frac{5}{2})$ & 0.011 &
 $\pi(6i \frac{13}{2}\frac{5}{2}) \to \nu(6i \frac{11}{2}\frac{5}{2})$ & 0.031 &
 $\pi(6i \frac{13}{2}\frac{5}{2}) \to \nu(6i \frac{11}{2}\frac{3}{2})$ & 0.008 \\
 &$\pi (6i \frac{13}{2}\frac{5}{2})\to \nu (6i\frac{11}{2}\frac{5}{2})$&0.013& &\\
 &$\pi (6i \frac{13}{2} \frac{5}{2})\to \nu (6i \frac{11}{2}\frac{3}{2})$&0.011& &\\
\hline
$^{238}$Pu
 & $\pi(6i \frac{11}{2}\frac{7}{2}) \to \nu(6i \frac{13}{2}\frac{7}{2})$ & 0.008 &
 $\pi(6i \frac{13}{2}\frac{1}{2}) \to \nu(6i \frac{11}{2}\frac{1}{2})$ & 0.010&
  $\pi(6i \frac{11}{2}\frac{7}{2}) \to \nu(6i \frac{13}{2}\frac{5}{2})$ & 0.005\\
 & $\pi(5f \frac{5}{2}\frac{3}{2}) \to \nu(5f \frac{5}{2}\frac{5}{2})$ & 0.009&&&\\
\hline
\end{tabular}
\caption{The same as in Table III but for the lower panels of Figs. 2-9.}
\end{table}
\begin{table}[!]
\begin{tabular}{|c||c|c|c|c|c|c|}
\hline 
Nucleus & \multicolumn{2}{c|}{1st peak} & \multicolumn{2}{c|}{2nd peak} & 
\multicolumn{2}{c|}{3rd peak} \\ \cline{2-7}
 & pnQRPA energy & Strength & pnQRPA energy & Strength & pnQRPA energy & Strength \\ \hline \hline
$^{76}$Se
 & 2.684 & 0.313 &
 4.904 & 0.116 &
 13.028 & 0.013 \\
 & & &6.169 &0.027& & \\
 & & &6.627 &0.021& & \\
\hline
$^{82}$Kr
 & 2.158 & 0.135 &
 5.313 & 0.118 &
 8.172 & 0.019 \\
\hline
$^{148}$Sm
 &2.196 & 0.139 &6.209&0.037&10.029 &0.011\\
 & 2.301 & 0.151 &7.408&0.017&11.899&0.012 \\
 &2.427&0.165     &    &     &      &\\
\hline
$^{150}$Sm
 & 3.051& 0.227 & 6.420
& 0.012 & &  \\
&3.102&0.134&7.369&0.018&&\\
&3.419&0.103&    &   & &\\
\hline
$^{154}$Gd
 & 2.071 & 0.081 & 3.169 & 0.061 &3.928 &0.039 \\
 & 2.491 & 0.140 & 3.600&0.046 & 4.170 &0.019 \\
 & & &  &  & 4.401 & 0.016 \\
\hline
$^{160}$Dy
 & 4.587 & 0.05 &
 5.616& 0.067 & 6.209&  0.063\\
 & 4.756& 0.076 &5.742 &0.113 &7.469 &0.051\\
 & 5.392& 0.365 &5.830& 0.059 &7.559& 0.054\\
 &      &       &5.968 &0.072 &     &       \\
\hline
$^{232}$U
 & 3.107& 0.011& 9.531 & 0.031& 16.069 & 0.008\\
 &4.917 & 0.013 &       &      &        &       \\
 &6.656 & 0.011 &       &      &        &        \\
\hline
$^{238}$Pu
 & 3.378 & 0.008 &
 11.751 & 0.010&17.689 &0.005  \\
 & 3.415 & 0.009 & & &\\
\hline
\end{tabular}
\caption{The same as in Table V, but for the lower panels of Figs. 2-9.}
\end{table}
\clearpage
 Of course, this feature is mainly caused by the fact that while for
Ge and Se transitions the states contributing to the GT resonance have an
energy separated by a gap from the upper two qp dipole configurations, in the heavier
nuclei such energy gap does not exist due to both
deformation effect on single particle energies and the fact that the last filled state is
far away from a major shell closure.

In Tables III and IV we list the single particle $\beta^-$ transitions characterized by the fact that
the corresponding two quasiparticle energy is the closest one to a pnQRPA state
contributing to the n-th peak with the strength given at its right side.
The pnQRPA energies for the states bringing the strength listed in Tables III
and IV are given in Tables V and VI, respectively.

The single particle transitions $\nu(nljI)\to \pi(nlj'I')$ which coherently contribute
to the collective transition $0^+\to 1^+$,  are characterized by the change of quantum
numbers j and I by at most one unit i.e.
$|\Delta j|$=0,1 and $|\Delta I|$=0,1. The  ($\Delta j, \Delta I$) values
for the single particle transition which represents the dominant component of the
pnQRPA state which carries the maximal strength in a GT resonance are
(1,1) ($^{76}$Ge), (0,1)($^{82}$Se), (0,1) ($^{148}$Nd), (0,1) ($^{150}$Nd),
(1,0) ($^{154}$Sm), (1,1) ($^{160}$Gd), (1,1) ($^{232}$Th), (0,1)($^{238}$U).
From Tables III and IV, it results that the GT resonances are admixtures of
$\Delta I+\Delta j=1,2$ transitions.

A common feature for all nuclei considered in the present paper is that the dominant
component of the pnQRPA state, i.e. that component which is excited with the largest
probability  by the GT transition operator, involves single particle states with
small $I$. Indeed, such  transitions $\nu I\to \pi I'$ have either $I$ or $I'$ equal to
$\frac{1}{2}$ or $\frac{3}{2}$. In the cases of $^{154}$Sm the angular momenta,
 involved
in the transition are equal to each other. The common value is
$\frac{3}{2}$. In Figs. 2-9 we give also the folded
2qp strength. One notes that the pnQRPA correlations push the strength to the
higher
energy. One of the main effects, comparing it with the 2qp picture, is that
it concentrates most of the strength in the GT resonance which is the most collective
pn excitation in the intermediate odd-odd nucleus.

Now, let us focus our attention on the $\beta^+$ strength distribution in the daughter
nuclei. These strengths are much weaker in magnitude than those characterizing
the $\beta^-$ strength in the mother nuclei. Another difference between the two
processes is that for the single beta minus process the maximum
strength is concentrated at relatively high energy, around the Gamow-Teller
resonance,
while in the beta plus decay, most of the strength lies around 5 MeV. Indeed, from the lower panels of Figs. 2-9, one sees that the first peak,\
is the highest one. Exception from this rule is $^{232}$Th where the $\beta^+$ strength is
quite small and its distribution has a peak lying around the energy of 10 MeV.
Switching on the QRPA correlations one notices a decrease of the 2qp strength.
Actually
the difference in strengths which appear for the peaks is distributed among the remaining
pnQRPA states, the total amount of strength in the two pictures being the same.
The two qp configurations which contribute most to the
first peaks shown in Figs. 2-9, are listed in Table VII. They are the dominant
components
of pnQRPA states with the energies given in Table VIII.
As in the case of the
single $\beta^-$ decay of the mother nuclei, here also most of the
dominant transitions take place
between states of low angular momenta ($I=\frac{3}{2},\frac{1}{2}$).
However, due to the small magnitude of the transition strengths here one
notices
several transitions between states with angular momenta equal to
$\frac{5}{2},\; \frac{7}{2},\; \frac{9}{2}$.

Inspecting the expression of the double beta transition amplitude, one
notes that the numerator of a chosen term from the sum, has three factors: i) one which determines the strength
of the $\beta^-$ transition to a particular state $|1_k\rangle _i$, ii) one whose hermitian
conjugate matrix element describes the $\beta^+$ transition to a state $|1_{k'}\rangle_f$ and iii) the
overlap of the states reached by the decay of the initial and final nuclei, respectively.
One expects that the maximum contribution to the GT transition amplitude $M_{GT}$
is achieved when the two  single beta matrix elements are maximal and moreover
the overlaps of the dipole states in the odd-odd system is maximum.
Therefore, we could ask ourselves whether among the peaks in the upper panels and those of the
lower panels there are pairs of peaks determined by states of maximal overlap.
From Tables III and VII one could identify many such pairs of peaks from the beta minus and beta plus
strength distributions.
For illustration we mention only one example. In $^{148}$Nd the maximum contribution
to the GT resonance is brought by the pnQRPA state of energy equal to 12.269MeV.
Indeed, the corresponding strength is 12.437 and moreover this is the leading strength.
The dominant amplitude for this state corresponds to the  single particle transition
$\nu(4g\frac{9}{2}\frac{5}{2})\to \pi(4g\frac{7}{2}\frac{3}{2})$.
On the other hand the strength distribution for $^{148}$Sm shows a third peak
determined by the pnQRPA state of energy equal to 11.899 MeV. Since the dominant
two quasiparticle component of this collective state, corresponds to the single particle
transition $\pi(4g\frac{7}{2}\frac{3}{2})\to \nu(4g\frac{9}{2}\frac{5}{2}$), as shown in
Table VII, one expects that the overlap of the pnQRPA states mentioned above
is maximally large.

The single beta decays strengths of a given nucleus satisfy the N-Z sum rule,
known
under the name of Ikeda sum rule \cite{Ike}. Our predictions for $\beta^-$ and
$\beta^+$ strengths
satisfy the Ikeda sum rule in the heavy isotopes while for $^{76}$Ge and $^{82}$Se
small deviations of 3 and 1.7\% respectively, are registered.

Let us analyze now the results for the double beta process, given in Table II.
The dipole-dipole interaction strengths have been chosen as given by the empirical formula
(\ref{chi}). The ratio of the $pp$ and $ph$ interaction strengths determines the $g_{pp}$
factor. As we already mentioned before, this A dependence for the interaction strengths
depends on single particle basis as well as on the truncation of the single particle space and
therefore its validity for the present formalism is questionable. As a matter of fact
for the lightest nuclei a larger value for $\chi$ approaches better the experimental situation while for
$^{238}$U a smaller value is more suitable.
Definitely, the safe way of fixing the ph and pp interaction strengths would be
to fit the position of GT resonance centroid of the odd-odd intermediate nucleus
and the half-lives of the $\beta^+$
decay of the of unstable nuclei in this mass region, respectively.
However since for the cases considered here the experimental data mentioned above
are lacking
, we adopted the empirical formula (\ref{chi}) just to obtain some reference results to be
compared with the ones obtained with the same interaction but different single
particle basis.

The denominator from  the equation defining M$_{GT}$ involves the $Q_{\beta\beta}$
values and the experimental energy of the first $1^+$. These values are given
in Table IX.
\begin{table}[h!]
\begin{tabular}{|c|cccccccc|}
\hline
Nucleus &$\hskip0.5cm$ $^{76}$Ge$\hskip0.5cm$ &$\hskip0.5cm$ $^{82}$Se
$\hskip0.5cm$ &$\hskip0.5cm$ $^{148}$Nd&$\hskip0.5cm$ $^{150}$Nd$\hskip0.5cm$ &
$\hskip0.5cm$ $^{154}$Sm$\hskip0.5cm$ &$\hskip0.5cm$ $^{160}$Gd$\hskip0.5cm$ &
      $^{232}$Th$\hskip0.5cm$ &$\hskip0.5cm$ $^{238}$U\\
\hline
Q$_{\beta\beta}$[MeV]&2.04 &3.01 &  1.93 &3.37 & 1.25& 1.73&0.85& 1.15\\
\hline
Nucleus &$^{76}$As &$^{82}$Br &$^{148}$Pm &$^{150}$Pm &$^{154}$Eu & $^{160}$Tb &
   $^{232}$Pa &$^{238}$Np \\
   \hline
E$_{1^+}$[keV]&44 & 75 & 137 & 137$^*$& 72 & 139 & 1000$^*$&244\\
\hline
\end{tabular}
\caption{The Q$_{\beta\beta}$-values for mother nuclei are given in units of MeV.
In the lowest row, the experimental energy for the first 1$^+$ states in the intermediate nuclei are given in
units of keV.$^{*)}$ For $^{150}$Pm and $^{232}$Pa there are not available data.
Therefore we take the ad hoc values characterized by an asterisk. Actually changing E$_{1^+}$ within 1 MeV
does not modify the order of magnitude of T$_{1/2}$}.
\end{table}        
Except for the case of ($^{76}$Ge; $^{76}$Se) all other pairs of (mother;daughter)
nuclei are characterized by only slightly different nuclear deformations. For
this reason
in our calculations the nuclear deformations of mother and daughter nuclei have been
considered equal to each other. The results for M$_{GT}$ and T$_{1/2}$ are listed in Table II.
They are compared with the experimental available data as well as with the
predictions of
those of Ref.\cite{Klap4}. Table II shows a good agreement between the
predicted T$_{1/2}$ for $^{82}$Se and $^{238}$U, and the corresponding
experimental half life given in Ref.\cite{Elli,Man,Kir} and Ref.\cite{Tret,Bara,Elli}, respectively. Our prediction for the half life of
$^{150}$Nd is 69 times lower than the corresponding lower experimental limit.
As shown in the second row for this nucleus, this discrepancy can be recovered by
changing
g$_{pp}$ to a value equal to 1.50. In this context we  would like to mention
that while for the lightest two nuclei from Table II, the M$_{GT}$ function of
g$_{pp}$ shows a very abrupt decreasing part, for the heavier nuclei the
cancellation point is reached with  a curve of a moderate slope. In the case of $^{150}$Nd
the cancellation value of g$_{pp}$ is larger than 1.8 and therefore the adjusted
value of
1.5 is still far away from the critical point where the pnQRPA breaks down.

The predicted half life of $^{76}$Ge shown  in the first row of Table II is only slightly
smaller than the lower limit of the corresponding experimental data. For this case, however,
the deformations for mother and daughter given in Ref. [40] are quite different
from each other. This feature challenged us to consider in our calculations different deformations for
$^{76}$Ge and $^{76}$Se. Therefore, we repeated the calculations for the decay
of $^{76}$Ge with the deformation $d_m=1.6$ for the mother and $d_d=1.9$ for the
daughter nucleus.
The pairing strengths for the new value for the
 nuclear deformation acquired by the mother nucleus are $G_p$=0.290, $G_n$=0.280.
 The results are shown in the second row of Table II.
 From there one remarks a good agreement with the experimental data for
 $T_{1/2}$.

To conclude, considering different deformations for mother and daughter nuclei
decreases the overlap matrix elements involved in $M_{GT}$. Due to this effect the
$T_{1/2}$ value is increased. It is an open question whether considering different deformations for
$^{150}$Nd and $^{150}$Sm would wash out the big discrepancy with the experimental data
shown in the first row for the decay of $^{150}$Nd. At the first glance one may
 say that
in order to have a positive answer to the question formulated above,
one needs a larger difference between
the two deformations than indicated in Refs. [39,40]. Since the half life is
sensitive to
the pairing properties one may suspect that for this case the proton-neutron
pairing might play an important role.

Comparing the results for $T_{1/2}$ obtained with our formalism with those obtained
in Ref.\cite{Klap4} by a different method one notices that for four emitters,
$^{148}$Nd,$^{150}$Nd, $^{154}$Sm, $^{160}$Gd, our predicted half lives are shorter
than in the above quoted reference while for the remaining nuclei the ordering
of the half lives is opposite.  In some cases the difference between the two
sets of predictions are in the range of two orders of magnitudes.
 Since the two methods are based on different
approaches for the transition amplitudes, it is an open question whether
these big discrepancies could get a consistent justification.

Finally we addressed the question of how the GT transition amplitude depends on
$g_{pp}$ and whether this dependence is influenced by the nuclear deformations.
The results of our investigation are presented in Figs. 12, 13. The input parameters of
single particle states and pairing interactions
are those of $^{76}$Ge. From Fig.12 one sees that for small
values of $g_{pp}$($\le$ 0.5), M$_{GT}$ depends monotonically on d while for
$g_{pp}\geq 0.5$
this property is lost. The repulsive character of the pp interaction causes the cancellation of M$_{GT}$
for a g$_{pp}$ around 3. As seen from Fig.12 the cancellation point depends on deformation.
Also the curves look of Fig 12 is not changed, the value of g$_{pp}$ where M$_{GT}$ is canceled
is quenched by the factor by which the strength $\chi$ is increased when one passes from Fig.12 to
Fig.13. In this context we recall that for spherical nuclei, the cancellation,
corresponding to the value of $\chi$ which reproduces the position of the GT resonance,
takes place for $g_{pp}\approx 1$. For this value the relation between the matrix elements
of the ${\it ph}$ and ${\it pp}$
two body interactions is given by the Pandya transformation. From Fig. 13 it results
that the cancellation points depend, as we already said, on deformation. It is an open question
whether the deformation dependence of the GT resonance is such that the cancellation point
of $M_{GT}$ is always brought to about 1.

\section{Summary and Conclusions} 
\label{sec:level5}
The main results described in the previous sections can be summarized as follows.
The two neutrino double beta decay transition amplitudes and half lives for eight isotopes have been
calculated within a pnQRPA approach based on a projected spherical single
particle basis.

The single particle energies are approximated by averaging a particle-core
Hamiltonian on the projected basis. Due to the fact that the core volume
conservation is properly taken into account, the resulting energies depend on
deformation in a similar manner as Nilsson levels. This feature suggests that
the results for two neutrino double beta decay rate provided by a pnQRPA formalism with such a basis
will be essentially different than those obtained in Ref.\cite{Rad2} where single particle
energies depend linearly on deformation.

First we studied the $\beta^-$ and $\beta^+$ strength distributions for mother and daughter
nuclei respectively. Both types of strengths are fragmented due to the nuclear deformation.
The position as well as the width of the GT resonance depend on nuclear mass.
Moreover,
while the GT resonance lies in the upper part of the pnQRPA energy spectrum
(the meaning of this statement is that beyond the GT resonance there is only a
little strength left to be distributed) the highest peak in the folded
strength distribution plot for the $\beta^+$ decay
appears always for low energies. This feature suggests that the GT resonance
is mainly influenced
by the ${\it ph}$ while the peak in the $\beta^+$ strength distribution, by
the ${\it pp}$
channels of the dipole-dipole
two-body interaction.
It seems that there is  a correspondence between the pnQRPA states of mother and daughter nuclei
contributing most to the folded strength distributions. The associated states,
due to the correspondence mentioned above, have maximal overlap and therefore
give the main contribution to the M$_{GT}$ value.
From Figs. 2-9 one remarks that the GT resonance strength depend on the atomic mass.
The larger  A, the larger the height of the resonance. Moreover the energy of the
resonance center is also an increasing function of A. For two  emitters,
$^{82}$Se and $^{154}$Sm, the GT resonance has a doublet structure.
This reminds us of the doublet structure of the dipole giant charge preserving
 resonances due to the coupling to the quadrupole degrees of freedom. Actually
 in this case also the doublet structure is a deformation effect and by this an effect caused by
 the quadrupole coordinates of the core.

The M$_{GT}$ and T$_{1/2}$ values have been first calculated by considering
equal deformations for mother and
daughter nuclei. The A dependence of the $ph$ and $pp$  proton-neutron dipole
interaction is taken as in Ref.\cite{Hom}.
The agreement with experimental data concerning the T$_{1/2}$
value of $^{82}$Se and
$^{238}$U
is very good good. The result for $^{76}$Ge is slightly smaller than the experimental data.
The discrepancy was removed by considering the deformation for $^{76}$Ge different
from that of $^{76}$Se. Indeed, this is the only case where according to Refs.[30,40]
the deformations for the two nuclei involved in the double beta decay are
quite different.
To bring the theoretical value of $T_{1/2}$ for $^{150}$Nd in agreement with
the experimental data
one needs a larger deformation difference than given in literature [39,40]. Moreover,
the pairing strength should deviate very much from what the difference of neighboring
even-even isotopes masses requires. Due to this feature for this case we
reproduced the experimental
half life by changing $g_{pp}$ from 0.11154 to 1.5. Note that the critical value of $g_{pp}$
for this isotope is 1.8.
It is noteworthy that for isotopes, $^{76}$Ge, $^{82}$Se, $^{238}$U, where the
calculated half lives agree with the corresponding experimental data the values used for $g_{pp}$
are small which results in having a small effect coming from the pp interaction
on this observable. As shown in Table II for $^{238}$U, cancelling $g_{pp}$ does
not alter the agreement with the experimental data. Then it arises the question
whether the $\it {pp}$ interaction is really needed at all in order to describe quantitatively
the double beta decay process. Is the large sensitivity of the single $\beta^+$
matrix elements, pointed out for the first time by Cha in Ref.\cite{Cha},
 a real effect
or just an artifact caused by the instability of the pnQRPA ground state
\cite{Rad1}? As shown in Table II for $^{76}$Ge, taking different deformations
for mother and daughter nuclei
brings an important effect on T$_{1/2}$ but not a dramatic change as claimed in Ref.
\cite{Pace}. The difference between the two descriptions consists in the fact
that here
the overlap matrix elements of the states in the mother and daughter nuclei
are estimated in a manner consistent with the pnQRPA approach, while in the quoted
reference the phonon operators are dissociated  and the overlaps are calculated
within the BCS and particle representations. Of course, in the latter case it is
not possible to get a real hierarchy of the effects involved.

As we stressed in Ref.\cite{Rad7}, going beyond pnQRPA approach, some forbidden
processes  might become possible. As an example, we studied the double beta
transition on excited collective states. In the near future we shall
investigate whether increasing the deformation for the daughter nuclei would substantially
increase the reduced decay probability for such processes.

\section{Appendix A}
\renewcommand{\theequation}{A.\arabic{equation}}
\setcounter{equation}{0}
\label{sec:level A}
In order to calculate the overlap matrix which enter the $M_{GT}$ expression,
we have to express the phonon operator for the mother nucleus in terms of the
phonon operator of the daughter nucleus, following the boson expansion procedure
\cite{Rad7}.
\begin{equation}
\Gamma^{\dagger}_{1\mu}(i,k)=\sum_{k'}[W^{f}_{i}(k,k')\Gamma^{\dagger}_{1\mu}(f,k')+Z^{f}_{i}(k,k')\Gamma_{1,-\mu}(f,k')(-)^{1-\mu}].
\label{Gamapl}
\end{equation}
where the amplitudes W and Z can be easily calculated as follows:
\begin{eqnarray}
W^{f}_{i}(k,k')&=&_i\langle 0|[\Gamma_{1\mu}(f,k'),\Gamma^{\dagger}_{1\mu}(i,k)]|0\rangle_f,\nonumber\\
Z^{f}_{i}(k,k')&=&_i\langle 0|[\Gamma^{\dagger}_{1\mu}(f,k),\Gamma^{\dagger}_{1,-\mu}(i,k')(-)^{1-\mu}]|0\rangle_f.
\label{WandZ}
\end{eqnarray}
It is clear that once these amplitudes are calculated, the overlap matrix elements are readily obtained:
\begin{eqnarray}
_i\langle 1_k|1_k'\rangle_f=W^{f}_{i}(k,k')_i\langle 0|0\rangle_f
\label{ovkk}
\end{eqnarray}
provided that the overlap of the two vacua is known.
In what follows we shall describe the necessary steps to derive the expressions of the two factors from the right hand side
of the above expression. By a direct calculation one finds:

\begin{eqnarray}
 W^{f}_{i}(k,k')^* & = & 
(X_k(i;pn) X_{k'}(f;p'n') - Y_k(i;pn) Y_{k'}(f;p'n')) 
 _i\langle 0|[A_{1\mu}(i;pn), A^\dagger_{1\mu'}(f;p'n')]|0\rangle _f  \nonumber \\
 &+& (X_k(i;pn) Y_{k'}(f;p'n') - Y_k(i;pn) X_{k'}(f;p'n')) 
 _i\langle 0|[(-)^{1-\mu}A^\dagger_{1-\mu}(i;pn),A^\dagger_{1\mu'}(f;p'n')]|0
\rangle _f\nonumber\\
\label{W*}
\end{eqnarray}
The symbol ``*'' stands for the complex conjugation operation.
The matrix elements of the commutators of the two quasiparticle operators are 
expressed further in terms of the anti-commutator of the single particle 
operators which is calculated as explained in Ref.\cite{Rad4}:
\begin{eqnarray}
\{ c_{\alpha I M}(i), c^\dagger_{\alpha' I' M'}(f) \} & = & {\cal N}^I_{nlj}(i) {\cal N}^{I'}_{n'l'j'}(f)
\sum_J (C^{1JI}_{I0I})^2 (N^{(g)}_J(d))^{-1} (N^{(g)}_J(d'))^{-1} \times \nonumber \\ 
 & & \mbox{} \times O^{(c)}_J(i,f) \delta_{II'} \delta_{jj'} \delta_{MM'} \equiv
 O^{\alpha I}_{if} \delta_{II'}\delta_{jj'}\delta_{\alpha\alpha'}\delta_{MM'}
\label{ccpl}
\end{eqnarray}
Here $N^{(g)}_{J}$ denotes the norm of the core projected state:

\begin{equation}
\varphi^{(g)}_{JM} (d)  =  N^{(g)}_J(d) P^J_{M0} e^{d(b^\dagger_{20} - b_{20})}
 |0\rangle _b; ; \\
\label{phigJM}
\end{equation}
where $|0\rangle_b$ denotes the vacuum state for the quadrupole bosons.
The overlap matrix for the  initial (i) and final (f) core states is denoted
by $ O^{(c)}_J (i,f)$,
and has the expression:

\begin{equation}
O^{(c)}_J (i,f) \equiv \langle \varphi_{JM}(d) | \varphi_{JM}(d') \rangle =
N^{(g)}_J (d) N^{(g)}_J (d') e^{-\frac{d^2+d'^2}{2}} (2L +1) I^{(0)}_J 
(\sqrt{dd'}).
\label{OcJ}
\end{equation}
where the factor $I^{(0)}_J$is defined by Eq.(\ref{IkJ}).
 The initial nucleus
 deformation is denoted by $d$ while $d'$ stands for the deformation parameter
 characterizing the final nucleus. Note that in the limit $d'\to d$ we have
 \begin{equation}
 O^{(c)}_J(i,f)\to 1,\;,\;O^{\alpha I}_{if}\to 1.
 \end{equation}

With these details, one further obtains:

\begin{equation}
_i\langle 0|[ A_{1\mu}(i;pn),A^\dagger_{1\mu'}(f;p'n') ]|0\rangle_f  =  \delta_{I_p I_{p'}} \delta_{I_n I_{n'}}
\delta_{\mu\mu'} (U^m_p U^d_p + V^m_p V^d_p)(U^m_n U^d_n + V^m_n V^d_n) 
 O^{\alpha_pI_p}_{if}O^{\alpha_nI_n}_{if},
\label{AiAfpl}
\end{equation}

and 

\begin{equation}
_i\langle 0|[(-)^{1-\mu} A^{\dagger}_{1,-\mu}(i;pn),A^\dagger_{1\mu'}(f;p'n') ]
|0\rangle_f  =  \delta_{I_p I_{p'}} \delta_{I_n I_{n'}}
\delta_{\mu\mu'} (U^m_p V^d_p - V^m_p U^d_p)(U^m_n V^d_n - V^m_n U^d_n) 
 O^{\alpha_pI_p}_{if}O^{\alpha_nI_n}_{if},
\label{AplApl}
\end{equation}

Thus, the final expressions for the amplitudes W and Z are:

\begin{eqnarray}
W^{f}_{i}(k,k')&=&\sum_{pn}\left[(X_k(i;pn) X_{k'}(f;pn) - Y_k(i;pn) 
Y_{k'}(f;pn))(U^i_p U^f_p + V^i_p V^f_p)(U^i_n U^f_n + V^i_n V^f_n)
\right. \nonumber\\
&+&\left.(X_k(i;pn) Y_{k'}(f;pn) - Y_k(i;pn) X_{k'}(f;pn)) (U^i_p V^f_p - V^i_p 
U^f_p)(U^i_n V^f_n - V^i_n U^f_n)
\right]O^{\alpha_pI_p}_{if}O^{\alpha_nI_n}_{if}  \nonumber\\
Z^{f}_{i}(k,k')&=&\sum_{pn}\left[(X_k(i;pn) X_{k'}(f;pn) - Y_k(i;pn) 
Y_{k'}(f;pn))(U^i_p V^f_p - V^i_p U^f_p)(U^i_n V^f_n - V^i_n U^f_n)
\right. \\
&+&\left.(X_k(i;pn) Y_{k'}(f;pn) - Y_k(i;pn) X_{k'}(f;pn)) (U^i_p U^f_p + V^i_p V^f_p)(U^i_n U^f_n +V^i_n V^f_n)
\right]O^{\alpha_pI_p}_{if}O^{\alpha_nI_n}_{if}\nonumber.
\label{Wif}
\end{eqnarray}

As for the ground states overlap, the result is
\begin{eqnarray}
_i\langle 0|0\rangle_f&\approx&_i\langle BCS|BCS\rangle_f \nonumber\\
&=&\prod_{p}\left[U^i_pU^f_p+V^i_pV^f_p\left(O^{\alpha_pI_p}_{if}\right)^2
\right]\prod_{n}\left[U^i_nU^f_n+V^i_nV^f_n\left(O^{\alpha_nI_n}_{if}\right)^2
\right].
\label{0i0f}
\end{eqnarray}

If the mother and daughter nuclei are characterized by the same nuclear
deformation, the corresponding overlap matrix elements are obtained from the
above formulae by replacing $d'$ by $d$. A good approximation of the resulting
equation is given by the expression:
\begin{equation}
_i\langle 1_k|1_{k'}\rangle _f=\sum_{pn}\left[X_{k}(i,pn)X_{k'}(f,pn)-Y_{k}(i,pn)Y_{k}(f,pn)\right].
\label{1k1k}
\end{equation}
The drawback of the procedure described above is that both factors of Eq.\ref{ovkk}
are evaluated within the framework of the BCS approximation while the matrix elements
describing the single $\beta^{\pm}$ transitions are calculated within the pnQRPA
approach. Moreover, due to the overlap factors $O^{\alpha I}_{if}$, even the
border of BCS frame
is crossed toward the particle representation. This inconsistency of the levels of approximations
makes the method doubtful, since it is not possible to define an hierarchy of various
effects\cite{Pace}. For example one could take care of a negligible contribution otherwise ignoring
an important one.
However, extending the spirit of the RPA approach to the case of different deformations for the initial
and final nuclei one obtains:

\begin{eqnarray}
W^{f}_{i}(k,k')&=&\sum_{pn}\left[X_k(i;pn) X_{k'}(f;pn) - Y_k(i;pn)
Y_{k'}(f;pn)\right], \nonumber \\
{_i}\langle 0|0\rangle {_f}& = & 1
\label{W00}
\end{eqnarray}
In this way the overlap matrix and the matrix elements characterizing the initial and
final nuclei are treated in an unitary fashion. The numerical calculations presented in the present paper
correspond to the overlap matrix determined by Eq.(\ref{W00}).
We postpone for a forthcoming paper, the description of the $\beta^{\pm}$ matrix elements
within a higher pnQRPA approach consistent with the procedure
presented in this Appendix for calculating the overlap matrix elements.

\begin{figure}[h]
\centerline{\psfig{figure=FFIG1.PS,width=8cm,bbllx=5cm,%
bblly=10cm,bburx=18cm,bbury=26cm,angle=0}}
\vskip8cm
\caption{(Color on line)Proton single particle energies for the N=3 and N=4 major shells,
given in units of $\hbar\omega_0$, are plotted as function of the deformation
parameter $d$. The quantum numbers on the right hand side are $nljI$ defined in Eqs. (2.5) and (2.7)}
\label{Fig. 1}
\end{figure}
\clearpage

\begin{figure}[h]
\centerline{\psfig{figure=FFIG2.PS,width=8cm,bbllx=5cm,%
bblly=10cm,bburx=18cm,bbury=26cm,angle=0}} 
\vskip8cm
\caption{(Color on line)Single $\beta^-$ strength, for $^{76}$Ge, and single $\beta^+$ strength
for $^{76}$Se, folded with a Gaussian function having the width of 1 MeV,
are plotted as a function of the energy within the BCS and pnQRPA approximation,
for three  values of the particle-hole interaction strength, $\chi$.The left and middle panels
correspond to d=1.9 while the right panel to $d_m=1.6$ and $d_d=1.9$.
For $\chi=0.35$ and $\chi=0.4$, we also give the experimental results from Ref.\cite{Mad,Hel}. }
\label{Fig. 2}
\end{figure}
\clearpage

\begin{figure}[h]
\centerline{\psfig{figure=FFIG3.PS,width=10cm,bbllx=5cm,%
bblly=10cm,bburx=18cm,bbury=26cm,angle=0}}
\vskip8cm
\caption{(Color on line)The same as in Fig. 2, but for the $\beta^-$ of $^{82}$Se and the
$\beta^+$ of $^{82}$Kr. Experimental data are from Refs\cite{Mad}. The right panels correspond
to d=0.2.}
\label{Fig. 3}
\end{figure}
\clearpage

\begin{figure}[h]
\centerline{\psfig{figure=FFIG4.PS,width=10cm,bbllx=5cm,%
bblly=10cm,bburx=18cm,bbury=26cm,angle=0}}
\vskip8cm
\caption{The same as in Fig. 2, but for the $\beta^-$ of $^{148}$Nd and the
$\beta^+$ of $^{148}$Sm.}
\label{Fig. 4}
\end{figure}
\clearpage

\begin{figure}[h]
\centerline{\psfig{figure=FFIG5.PS,width=8cm,bbllx=5cm,%
bblly=10cm,bburx=18cm,bbury=26cm,angle=0} }
\vskip8cm
\caption{The same as in Fig. 2, but for the $\beta^-$ of $^{150}$Nd and the
$\beta^+$ of $^{150}$Sm.
}
\label{Fig. 5}
\end{figure}
\clearpage

\begin{figure}[h]
\centerline{\psfig{figure=FFIG6.PS,width=8cm,bbllx=5cm,%
bblly=10cm,bburx=18cm,bbury=26cm,angle=0} }
\vskip8cm
\caption{The same as in Fig. 2, but for the $\beta^-$ of $^{154}$Sm and the
$\beta^+$ of $^{154}$Gd.}
\label{Fig. 6}
\end{figure}
\clearpage

\begin{figure}[h]
\centerline{\psfig{figure=FFIG7.PS,width=10cm,bbllx=5cm,%
bblly=10cm,bburx=18cm,bbury=26cm,angle=0} }
\vskip8cm
\caption{The same as in Fig. 2, but for the $\beta^-$ of $^{160}$Gd and the
$\beta^+$ of $^{160}$Dy.}
\label{Fig. 7}
\end{figure}
\clearpage

\begin{figure}[h]
\centerline{\psfig{figure=FFIG8.PS,width=8cm,bbllx=3cm,%
bblly=10cm,bburx=18cm,bbury=26cm,angle=0} }
\vskip9cm
\caption{The same as in Fig. 2, but for the $\beta^-$ of $^{232}$Th and the
$\beta^+$ of $^{232}$U.}
\label{Fig. 8}
\end{figure}
\clearpage

\begin{figure}[h]
\centerline{\psfig{figure=FFIG9.PS,width=8cm,bbllx=3cm,%
bblly=10cm,bburx=18cm,bbury=26cm,angle=0} }
\vskip9cm
\caption{The same as in Fig. 2, but for the $\beta^-$ of $^{238}$U and the
$\beta^+$ of $^{238}$Pu.}
\label{Fig. 9}
\end{figure}
\clearpage

\begin{figure}[h]
\centerline{\psfig{figure=FFIG10.PS,width=8cm,bbllx=3cm,%
bblly=10cm,bburx=18cm,bbury=26cm,angle=0} }
\vskip9cm
\caption{The single $\beta^-$, for $^{82}$Se, and single $\beta^+$, 
for $^{82}$Kr, are plotted as a function of the energy within the pnQRPA
approach, for two values of the particle-hole interaction strength, $\chi$.}
\label{Fig. 10}
\end{figure}
\clearpage

\begin{figure}[h]
\centerline{\epsfig{figure=FFIG11.PS,width=8cm,bbllx=3cm,%
bblly=10cm,bburx=18cm,bbury=26cm,angle=0}}
\vskip6cm
\caption{The single $\beta^-$, for $^{232}$Th, and single $\beta^+$, 
for $^{232}$U, are plotted as function of the energy within the pnQRPA
approach, for $\chi=0.11$.}
\label{Fig. 11}
\end{figure}
\clearpage

\begin{figure}[h]
\centerline{\epsfig{figure=FFIG12.PS,width=8cm,bbllx=3cm,%
bblly=10cm,bburx=18cm,bbury=26cm,angle=-90}}
\vskip3cm
\caption{(Color on line) The Gamow-Teller amplitude $M_{GT}$ for the transition
$2\nu\beta\beta$ is represented as a function of
$g_{pp}$, for a particular value of the particle-hole interaction
strength, $\chi=0.1$.  }
\label{Fig. 12}
\end{figure}
\clearpage
\begin{figure}[h]
\centerline{\epsfig{figure=FFIG13.PS,width=8cm,bbllx=3cm,%
bblly=10cm,bburx=18cm,bbury=26cm,angle=-90}}
\vskip2cm
\caption{(Color on line) The same as in Fig. 12 but for  $\chi=0.3$.  }
\label{Fig. 13}
\end{figure}
\clearpage

\end{document}